\documentclass[journal,12pt,onecolumn,draftclsnofoot]{IEEEtran}
\usepackage{cite}
\usepackage{amsfonts,amssymb,amsthm,amsbsy,amsmath,paralist,bm,ifthen,color}
\usepackage{algorithm,algorithmic}
\usepackage{graphicx}
\usepackage[font=footnotesize]{caption}
\newtheorem{remark}{Remark}
\newtheorem{proposition}{Proposition}
\usepackage{xcolor}
\usepackage{relsize}
\usepackage[numbers,sort&compress,square]{natbib}
\usepackage{array,booktabs,multirow}
\usepackage{setspace}
\usepackage[bookmarks=true,colorlinks=false,bookmarksopen=true,bookmarksnumbered=true,backref=true]{hyperref}

\usepackage{subfigure}
\usepackage{stfloats}
\newtheorem{Lemma}{Lemma}

\newcommand{\Hb}    {\ensuremath{\bf{H}}}

\newcommand{\xb}    {\ensuremath{\bf{x}}}

\newcommand{\Rdom}{\ensuremath{\mathbb{R}}}
\newcommand{\Cdom}{\ensuremath{\mathbb{C}}}

\begin{document}
	\title{Robust Symbol-Level Precoding and Passive Beamforming for IRS-Aided Communications}
	\author{Guangyang Zhang, Chao Shen, Bo Ai and Zhangdui Zhong\\
		{\small Beijing Jiaotong University, State Key Laboratory of Rail Traffic Control and Safety, Beijing, China 100044}\\
		\small Frontiers Science Center for Smart High-speed Railway System, Beijing, China 100044\\
		\small Beijing Engineering Research Center of High-speed Railway Broadband Mobile Communications, China 100044\\
		\small Key Laboratory of Railway Industry of Broadband Mobile Information Communications, Beijing, China 100044\\
		{\small Email: \{guangyangzhang, chaoshen, boai, zhdzhong\}@bjtu.edu.cn}
	}
	\maketitle
	
\begin{abstract}
This paper investigates a joint beamforming design in a multiuser multiple-input single-output (MISO) communication network aided with an intelligent reflecting surface (IRS) panel. The symbol-level precoding (SLP) is adopted to enhance the system performance by exploiting the multiuser interference (MUI) with consideration of bounded channel uncertainty. The joint beamforming design is formulated into a nonconvex worst-case robust programming to minimize the transmit power subject to single-to-noise ratio (SNR) requirements. To address the challenges due to the constant modulus and the coupling of the beamformers, we first study the single-user case. Specifically, we propose and compare two algorithms based on the semidefinite relaxation (SDR) and alternating optimization (AO) methods, respectively. It turns out that the AO-based algorithm has much lower computational complexity but with almost the same power to the SDR-based algorithm. Then, we apply the AO technique to the multiuser case and thereby develop an algorithm based on the proximal gradient descent (PGD) method. The algorithm can be generalized to the case of finite-resolution IRS and the scenario with direct links from the transmitter to the users. Numerical results show that the SLP can significantly improve the system performance. Meanwhile, 3-bit phase shifters can achieve near-optimal power performance.
\end{abstract}


\section{Introduction}
\IEEEPARstart{T}{he} future sixth generation (6G) wireless communication system is expected to achieve a goal of ubiquitous wireless intelligence in ten years.
The telepresence and mixed reality are widely accepted as typical services in the era of 6G.
To support these services, the key performance indicators (KPI) including the peak rate, connection density should be carefully evaluated.
For example, the peak data rate will be up to 100-1000 Gbps and the connection density should reach 100 devices per cubic meter \cite{latva2020key}.
Hence, several novel technologies are envisioned to provide ultra-high speed transmission with deep coverage of massive users.
Among them, the millimeter wave (mmWave) and terahertz (THz) communications are considered to be potential solutions to achieve a very high data rate with the band range from 30 GHz to 3 THz \cite{8663550,8387218,9269930}.
Although the ultra-narrow beams generated by large-scale arrays admit massive users, the path loss and the penetration loss are severe due to the inherent property of high-frequency wave propagation. Therefore, ensuring good-quality and seamless coverage in the mmWave and THz communications is challenging, especially for the scenarios with rich blockages.
Recently, the intelligent reflecting surface (IRS) is introduced to the 6G wireless communications, and it is promising to improve the spectrum and energy efficiency by establishing configurable wireless channels \cite{8936989,9113273,8910627,9110912}.

A fundamental scheme to harness the benefits of IRS-aided communication networks is to jointly design the transmit beamforming and passive beamforming vectors \cite{8811733,9105111,9246254}.
The main challenges of joint beamforming lie in the coupling between the beamformers and the constant-modulus constraints over the passive phase shifters.
To solve these challenges, the authors in \cite{8811733} jointly designed beamforming vectors by adopting the semidefinite relaxation (SDR) and alternating optimization (AO) methods in a multi-input single-output (MISO) communication system.
The work was extended to the case with multiple IRS panels in \cite{9105111}, where the weighted sum rate is maximized, and the constant modulus constraint is resolved by applying the Riemannian manifold conjugate gradient method.
Meanwhile, a multi-cell downlink MISO system aided with IRS was studied in \cite{9246254} without consideration of the constant modulus requirement.
It should be noticed that perfect channel state information (CSI) is assumed to be available at the transmitters for all these works.

Since the perfect CSI is of critical importance for IRS-aided wireless communication systems, the channel estimation is elaborately examined in \cite{9195133,9130088,9081935}.
In \cite{9195133}, the authors proposed two methods to estimate the channel in a multiuser orthogonal frequency division multiple access (OFDMA) system.
The authors in \cite{9130088} proposed a three-phase pilot-based channel estimation framework which exploits the correlations among the IRS reflected channels.
Besides, a joint training sequence and reflection pattern was studied under the criterion of minimum mean square error (MMSE) in \cite{9081935}.
However, the CSI estimation accuracy depends on the training overhead, and thus the CSI error is inevitable for IRS-aided systems with limited training pilots.
In view of this, some researches on robust beamforming were carried out in IRS-aided systems with consideration of imperfect CSI.
For the case of Gaussian CSI error,  \cite{9117093} considered an MMSE-optimal beamforming in a single-user MISO system,
while \cite{9148947} investigated a chance-constrained beamforming design in a multi-user MISO system.
Furthermore, for the case of bounded CSI error, \cite{9110587} studied a worst-case robust beamforming design, where the constant modulus constraint is treated by penalty convex-concave procedure \cite{lipp2016variations}.

The performance gain obtained by beamforming design in a multiuser system is mainly due to its interference mitigation capability.
One way to mitigate the multiuser interference (MUI) is to simply perform the block-level precoding (BLP).
However, the symbol-level precoding (SLP) can convert the harmful MUI into constructive interference (CI), and thus improve the symbol error rate (SER) performance \cite{7042789,7103338,8647896}.
Recently, the SLP technique is introduced to the IRS-aided wireless communications.
A power minimization problem with phase-shift keying (PSK) modulation and an SER minimization problem were studied in \cite{9219206} and \cite{9099879}, respectively.
However, it is difficult to extend the work  \cite{9219206} to the other modulation constellations.
Moreover, both works assumed that the perfect CSI is available at the transmitter.

In this paper, we investigate a joint robust beamforming design in an IRS-aided multi-user communication system, where the bounded CSI error is taken into consideration, and the symbol level precoding is adopted to exploit the MUI.
The goal of the design is to minimize the transmit power under a given signal-to-noise ratio (SNR) requirement for each user.
Two worst-case robust beamforming design algorithms are developed for the single-user and the multiuser scenarios.
The main contributions of this paper are as follows:
\begin{itemize}
	\item We formulate the multiuser joint beamforming design with bounded CSI error as a worst-case robust optimization problem, which is applicable to all digital modulation constellations, and the phase shifters with infinite and finite resolutions.
	
	\item We propose an SDR-based algorithm to tackle the single-user case, and an AO-based algorithm with the similar power performance is developed to reduce the computational complexity, which admits a closed-form solution at each iteration.
	 For the multiuser case, we apply the AO technique and thereby develop an algorithm based on the proximal gradient descent (PGD) method to resolve the constant modulus constraints.
	The proposed algorithm can be generalized to the cases of finite-resolution IRS and the scenario with direct links from the base station (BS) to the users.
	
	\item Numerical results show that the SLP can significantly improve the system performance, and the constellation has a significant impact on the transmit power to satisfy the network's requirement.
	Meanwhile, it is shown that 8 phase shift levels can achieve near-optimal performance in terms of power.
\end{itemize}

The rest of the paper is organized in the following way.
Section $\rm\uppercase\expandafter{\romannumeral 2}$ introduces the system model and provides the problem formulation in detail.
In Section $\rm\uppercase\expandafter{\romannumeral 3}$, two methods are proposed and analyzed for the single-user case.
In Section $\rm\uppercase\expandafter{\romannumeral 4}$, we propose an algorithm to solve the problem for the multiuser scenario.
Then, some simulation results are provided in Section $\rm\uppercase\expandafter{\romannumeral 5}$.
Finally, the conclusions are given in Section $\rm\uppercase\expandafter{\romannumeral 6}$.

\textit{Notation}:
The notations used in this paper are as follows. Plain lower case letters are used to denote scalars, e.g., $\phi$.
Boldface lower and upper letters are used to denote column vectors and matrices, respectively, e.g., $\mathbf{x}$ and $\mathbf{G}$.
An $n$-dimensional identity matrix is denoted by either  ${\bf I}_n$ or  ${\bf I}$ when the dimension is clear from the context.
For a complex scalar $a$, $\Re(a)$ and $\Im(a)$ denote the real and imaginary part of $a$, respectively, and $|a|$ denotes its magnitude. Additionally, $\angle a$ denotes the angle of $a$.
For a vector, $\|\cdot\|_2$ denotes its Euclidean norm, and let $(\cdot)^{T}$ and $(\cdot)^{H}$ denote the transpose and Hermitian operations, respectively. Besides, $\text{diag}(\mathbf{x})$ indicates the diagonal matrix whose diagonals are the elements of $\mathbf{x}$.
While $\Rdom^n$ and $\Cdom^n$ stand for the set of $n$-dimensional real and complex vectors.
$x \sim\mathcal{CN}(\mu,\sigma^2)$ means that $x$ is a circularly symmetric complex Gaussian (CSCG) random variable with mean $\nu$ and variance $\sigma^2$.
For a square matrix $\mathbf{S}$, let $\text{tr}(\mathbf{S})$ denote its trace, and $\mathbf{S}\succeq0$ means that $\mathbf{S}$ is a semidefinite matrix.
Particularly, $\mathbf{x}\succeq\mathbf{y}$ means that each element of $\mathbf{x}$ is equal or greater than its corresponding element of $\mathbf{y}$.

\section{System Model and Problem Formulation}
We consider a downlink communication system aided by the IRS, as illustrated in Fig. 1. It consists of a BS with $M$ antennas, $K$ single-antenna users, and an IRS panel with $N$ reflective elements. Besides, there is a controller between the BS and the IRS so that coordinated communication can be achieved. As shown in Fig. 1, due to the blockage between the BS and $K$ users, the desired signal for the users can only be reflected by the IRS panel. For each element of the IRS, it can introduce a phase shift to the incident signal but the amplitude of the signal cannot be changed. Specifically, let $\bm{\theta}=[e^{j\phi_1},\cdots,e^{j\phi_N}]^H\in\mathbb{C}^{N}$ denote the phase shift vector introduced by the IRS with $\phi_n\in [0,2\pi)$ for $n=1,\cdots,N$. Besides, we denote $\bm{{\rm G}}\in\mathbb{C}^{N\times M}$ and $\mathbf{h}_k\in\mathbb{C}^N, k=1,\cdots,K$ as the BS-IRS link and the channel from the IRS to the user $k$.
\begin{figure}
\begin{center}
\includegraphics[width=7.5 cm]{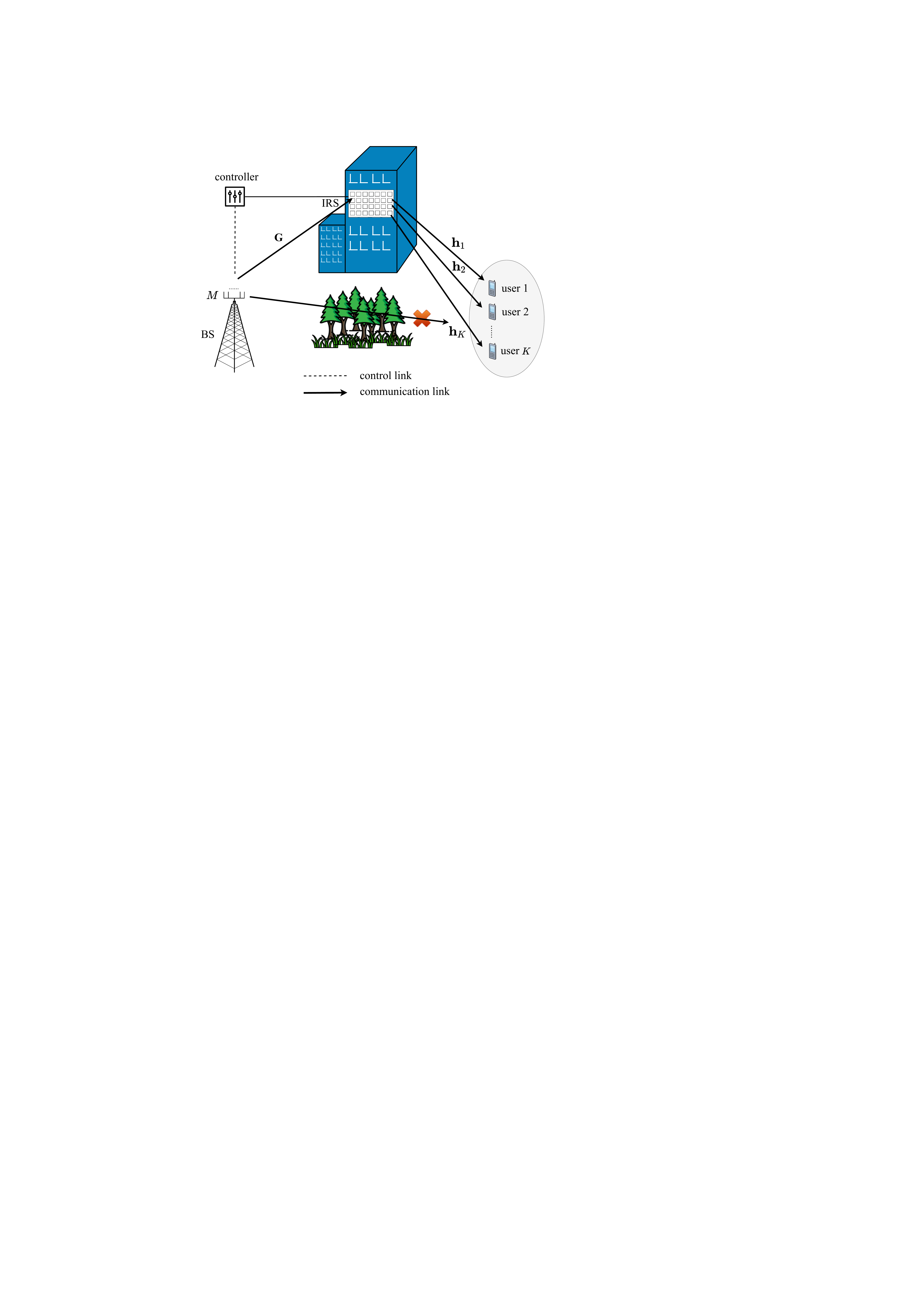}
\caption{An IRS-aided wireless communication system.}
\label{Casimir}
\end{center}
\vspace{-0.5cm}
\end{figure}

\subsection{Signal Model}
Different from the most of the IRS-related researches, we consider that the symbol-level precoding rather than the traditional block linear beamforming is performed at the BS. The desired symbols $\{s_k\}$ for the users in a symbol time are independent, and let $\bm{s}\triangleq[s_1,\cdots,s_K]^T\in\mathbb{C}^{K}$. Each symbol $s_k$ for the user $k$ is taken from a finite constellation point set independently. In the context of the symbol-level precoding, the original symbol vector $\bm{s}$ is mapped to a vector $\bm{{\rm x}}\in\mathbb{C}^{M}$ which is transmitted by $M$ antennas of the BS. Therefore, the received signal at the user $k$ can be written as
\begin{align}
y_k=\mathbf{h}_k^H\text{diag}(\bm{\theta}^H)\bm{{\rm G}}{\rm\mathbf{x}}+z_k=\bm{\theta}^H\bm{{\rm H}}_k{\rm\mathbf{x}}+z_k,
\end{align}
where $\bm{{\rm H}}_k\triangleq\text{diag}(\mathbf{h}_k^H)\bm{{\rm G}}\in\mathbb{C}^{N\times M}$ is referred to as the concatenated channel, and $z_k$ is the additive Gaussian noise with $z_k\thicksim\mathcal{CN}(0,\sigma^2_k)$. In order to simplify analysis below, we transform the complex system into a real system.
To this end, we define
\begin{subequations}
\begin{align}
\tilde{{\rm\mathbf{x}}}&\triangleq[\Re({\rm\mathbf{x}}^T), \Im({\rm\mathbf{x}}^T)]^T\in\mathbb{R}^{2M},\\
\tilde{{\rm\mathbf{y}}}_k&\triangleq[\Re(y_k), \Im(y_k)]^T\in\mathbb{R}^{2},\\
\tilde{{\rm\mathbf{z}}}_k&\triangleq[\Re(z_k), \Im(z_k)]^T\in\mathbb{R}^{2},\\
\tilde{\bm{{\rm H}}}_k&\triangleq\mathcal{T}(\bm{{\rm H}}_k)\in\mathbb{R}^{2N\times 2M},
\end{align}
\end{subequations}
where $\mathcal{T}(\cdot)$ is a transformation operator of a complex-valued matrix $\bm{{\rm U}}\in\mathbb{C}^{m\times n}$ and defined as
\begin{equation}
\mathcal{T}(\bm{{\rm U}})=
\left[
  \begin{array}{cc}
    \Re(\bm{{\rm U}}) & -\Im(\bm{{\rm U}}) \\
    \Im(\bm{{\rm U}}) & \Re(\bm{{\rm U}}) \\
  \end{array}
\right]\in\mathbb{R}^{2m\times 2n}.
\end{equation}

Hence, by splitting the complex-valued signal into real and imaginary parts, the received signal at the user $k$ can be expressed equivalently as
\begin{align}
\tilde{\mathbf{y}}_k=&\mathcal{T}(\bm{\theta}^H\bm{{\rm H}}_k)\tilde{{\rm\mathbf{x}}}+\tilde{{\rm\mathbf{z}}}_k=\mathbf{\Theta}\tilde{\bm{{\rm H}}}_k\tilde{{\rm\mathbf{x}}}+\tilde{{\rm\mathbf{z}}}_k,
\end{align}
where ${\rm\mathbf{\Theta}}\triangleq\mathcal{T}(\bm{\theta}^H)\in\mathbb{C}^{2\times 2N}$.

\subsection{Channel Uncertainty Model}
Obtaining an accurate channel estimation is challenging for any wireless communication system, especially for a system enhanced by a passive IRS panel. In this paper, we adopt the approach raised in \cite{9167248} to get the channel estimation. Its main idea is taking advantage of the orthogonality of the pilot sequences transmitted by the users. The protocol can be summarized as follows. During the $n$th time slot, all $K$ users transmit the orthogonal pilot sequences and the signal is received at the BS via the IRS reflection. Meanwhile, only the $n$th IRS element is working and other $N-1$ elements are turned off for $n=1,\cdots,N$. Thus, in the $n$th time slot, the signal $\mathbf{R}_n\in\mathbb{C}^{L\times M}$ received at the BS can be expressed as
\begin{align}
\mathbf{R}_n=\sum_{k=1}^Kh_{kn}^H\mathbf{p}_k\mathbf{g}_n^H+\mathbf{Z}_n, \forall n,
\end{align}
where $\mathbf{g}_n\in\mathbb{C}^{M}$ denotes the $n$th column of $\mathbf{G}^H$, $h_{kn}\in\mathbb{C}$ represents the $n$th element of the channel gain vector $\mathbf{h}_k$, $\mathbf{p}_k\in\mathbb{C}^{L}$ is the pilot sequence of the user $k$, and $\mathbf{Z}_n\in\mathbb{C}^{L\times M}$ is the additive Gaussian noise. Due to the orthogonality of the pilot sequences, the concatenated channel between the user $k$ and the BS via the $n$th IRS element can be estimated based on
\begin{align}
\mathbf{R}_n^H\mathbf{p}_k=h_{kn}\mathbf{g}_n+\mathbf{z}_{nk},
\end{align}
where $\mathbf{z}_{nk}=\mathbf{Z}_n^H\mathbf{p}_k\in\mathbb{C}^M, \forall n,k$. As a result, the estimation of the channel between the user $k$ and the BS can be written as
\begin{align}
\hat{\bm{{\rm H}}}_k=[\mathbf{R}_1^H\mathbf{p}_k,\cdots,\mathbf{R}_N^H\mathbf{p}_k]^H\in\mathbb{C}^{N\times M}.
\end{align}
Inevitably, there will exist an estimation error between the actual channel $\mathbf{H}_k$ and the channel estimation $\hat{\bm{{\rm H}}}_k$, which can be denoted by $\bm{\Delta}_k\triangleq\bm{{\rm H}}_k-\hat{\bm{{\rm H}}}_k$. Throughout this paper, we assume that the channel estimation error $\bm{\Delta}_k$ is Frobenius-norm bounded by
\begin{align}
\|\bm{\Delta}_k\|_F\leq\frac{\sqrt{2}}{2}\delta_k,~\forall k,
\end{align}
where $\delta_k>0$ is the radius of the estimation error. We believe that the proposed scheme in this paper can be easily extended to the case with elliptically bounded CSI errors, i.e., $\text{trace}(\bm{\Delta}_k\mathbf{P}_k\bm{\Delta}_k^H)\leq\delta_k^2$ for some positive semidefinite (PSD) matrix $\mathbf{P}_k$.

\subsection{Problem Formulation}
In this paper, we consider a robust beamforming design of the active and passive antennas at the BS and the IRS, respectively. The goal of the design is to minimize transmit power under the given SNR requirements for all $K$ users. Besides, the symbol-level precoding is adopted at the BS to improve the SER performance. For the symbol-level precoding, the noise-free received signal at the receiver is expected to locate in the constructive interference region (CIR) which corresponds to its desired symbol. Specifically, for a given constellation point, its CIR is established by the hyperplanes which separate it from its neighboring constellation points. Hence, the CIR is dependent on the location of the constellation point.

According to formula (4), the desired noise-free signal for the user $k$ can be expressed as $\mathbf{\Theta}\tilde{\bm{{\rm H}}}_k\tilde{\mathbf{x}}$. Thus, the CIR constraint for the user $k$ can be modeled as a general form as \cite{8299553}
\begin{align}\label{Formula1}
\mathbf{A}_k\mathbf{\Theta}\tilde{\bm{{\rm H}}}_k\tilde{\mathbf{x}}\succeq\sigma_k\sqrt{\gamma_k}\mathbf{b}_k,
\end{align}
where ${\rm \mathbf{A}}_k\in\mathbb{R}^{S_k\times 2}$ and ${\rm \mathbf{b}}_k\in\mathbb{R}^{S_k}$ determine the CIR of the desired constellation point for the user $k$, $S_k$ denotes the number of the CIR's boundaries, and $\gamma_k$ is the SNR requirement for the user $k$,

\begin{remark}
For the quadrature phase shift keying (QPSK) and 8-PSK constellations, the CIR of each constellation point is determined by two hyperplanes, as shown in Fig. 2(a) and (b). As an example, the CIR of the QPSK constellation point in the second quadrant can be expressed in a compact form as
\begin{align}
\mathcal{S}\triangleq\{\tilde{\mathbf{s}}|\mathbf{A}^{\rm QPSK}\tilde{\mathbf{s}}\succeq\mathbf{b}\},
\end{align}
where $\mathbf{A}^{\rm QPSK}=\begin{bmatrix}-\sqrt{2}&0\\0&\sqrt{2}\end{bmatrix}$ and $\mathbf{b}=[1, ~1]^T$. But, the number of the CIR's boundaries is not identical for all constellations. For the case of 16 quadrature amplitude modulation (16-QAM), as shown in Fig. 2 (c), the CIR of the red constellation point is established only by two hyperplanes; however, the CIRs of the blue and green constellation points are determined by 3 and 4 hyperplanes, respectively.
\end{remark}

\begin{figure*}[htbp]
\centering
\subfigure[QPSK]{
\begin{minipage}{0.3\linewidth}
\centering
\includegraphics[width=4.5 cm]{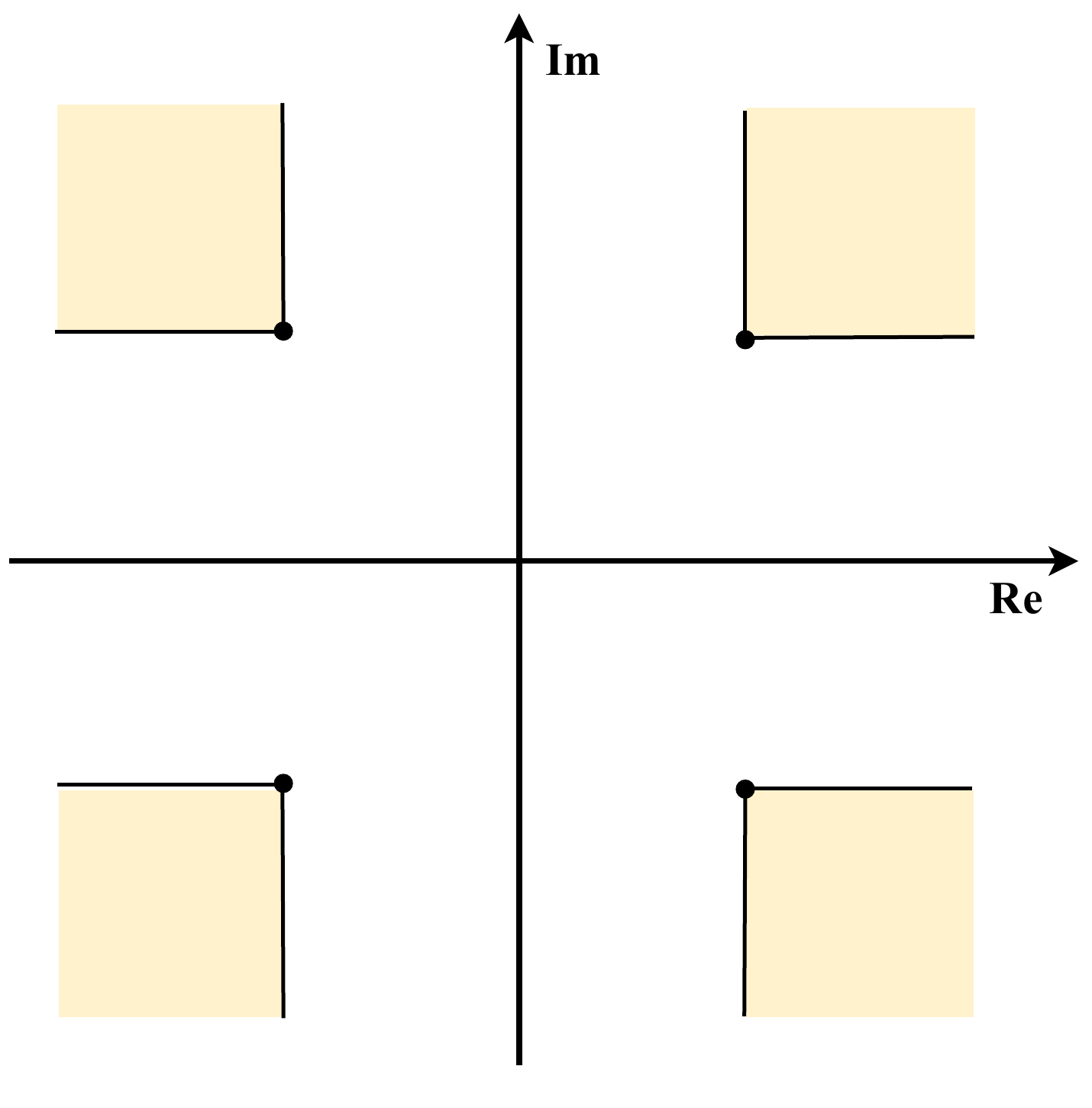}
\end{minipage}%
}
\subfigure[8-PSK]{
\begin{minipage}{0.3\linewidth}
\centering
\includegraphics[width=4.5 cm]{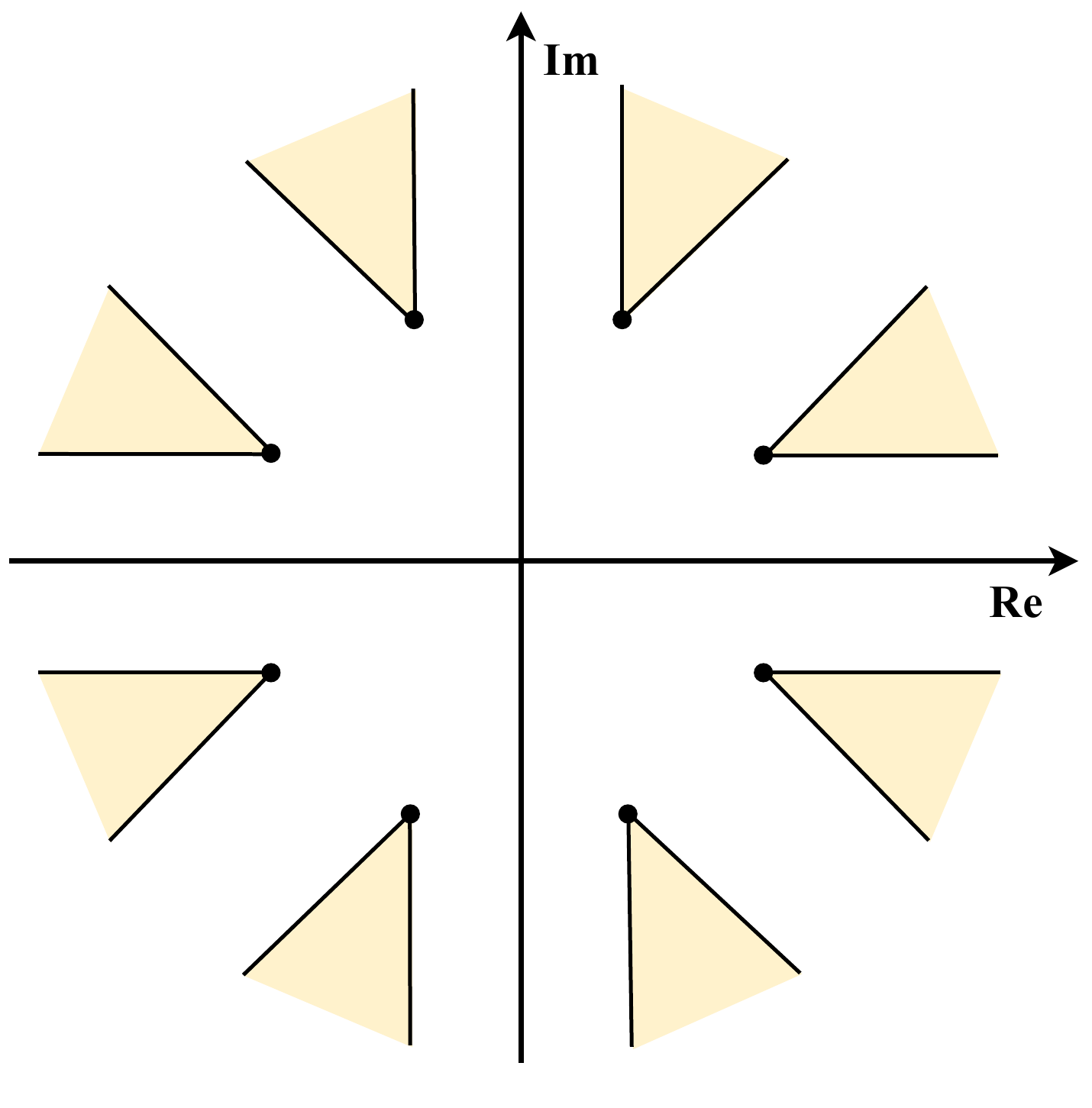}
\end{minipage}%
}
\subfigure[16-QAM]{
\begin{minipage}{0.3\linewidth}
\centering
\includegraphics[width=4.5 cm]{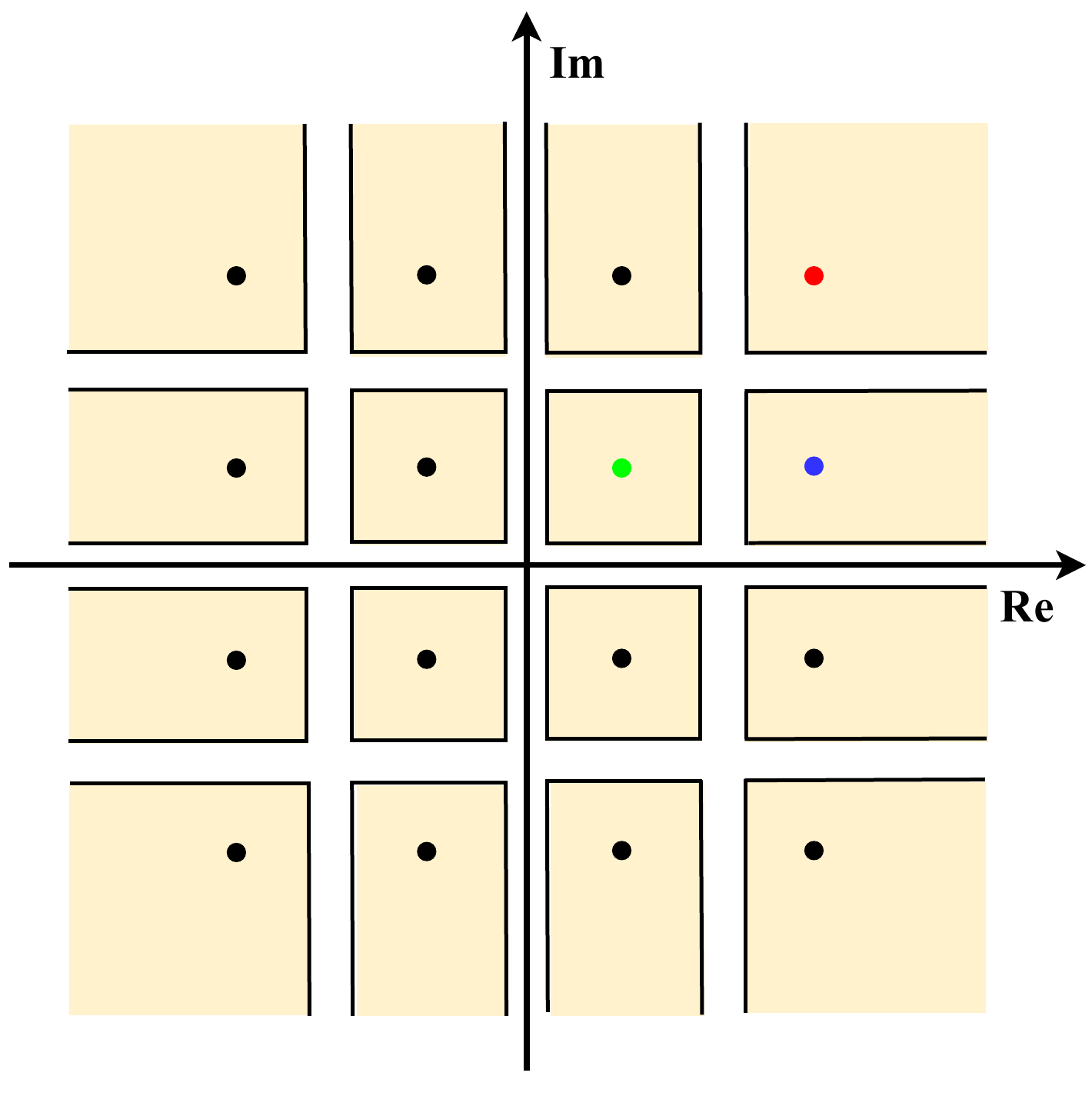}
\end{minipage}
}
\caption{Constellations and their CIRs which are denoted by the orange areas.}
\end{figure*}

For simplicity of analysis, by denoting ${\rm\mathbf{A}}_k=[{\bf a}_{k,1}, \cdots, \mathbf{a}_{k,S_k}]^T$ and ${\rm\bf{b}}_k=[b_{k,1}, \cdots, b_{k,S_k}]^T$, we can transform the constraint \eqref{Formula1} into the constraints as following
\begin{align}\label{Formula2}
{\bf{a}}_{k,i}^T\mathbf{\Theta}\tilde{\bm{{\rm H}}}_k\tilde{\mathbf{x}}\geq\sigma_k\sqrt{\gamma_k}b_{k,i}, ~i=1,\cdots,S_k.
\end{align}

To further reveal the hidden structure of \eqref{Formula2}, let us define $\tilde{\bm{\theta}}\triangleq[\Re\{\bm{\theta}^H\}, \Im\{\bm{\theta}^H\}]^T$. Then, by applying the property of the Kronecker product, we can obtain
\begin{align}
\mathbf{a}_{k,i}^T\mathbf{\Theta}=\tilde{\bm{\theta}}^T\bm{{\rm D}}_{k,i}, ~\forall i,
\end{align}
where $\bm{{\rm D}}_{k,i}\in\mathbb{R}^{2N\times 2N}$ is defined as
\begin{align}
\bm{{\rm D}}_{k,i}=(\mathbf{I}_2\otimes\mathbf{a}_{k,i}^T)
\begin{bmatrix}
1&0\\
0&1\\
0&-1\\
1&0
\end{bmatrix}\otimes\mathbf{I}_N.
\end{align}
Therefore, we can obtain alternate expressions of the constraints \eqref{Formula2}
\begin{align}
\tilde{\bm{\theta}}^T\bm{{\rm D}}_{k,i}\tilde{\bm{{\rm H}}}_k\tilde{\mathbf{x}}\geq\sigma_k\sqrt{\gamma_k}b_{k,i}, ~\forall i.
\end{align}

Recall that each IRS element only introduces a phase shift without changing the amplitude of the incident signal. Then, according to the definition of $\tilde{\bm{\theta}}$, we have $\tilde{\theta}_n^2+\tilde{\theta}_{n+N}^2=1, n=1,\cdots,N$, where $\tilde{\theta}_n$ denotes the $n$th element of $\tilde{\bm{\theta}}$. Consequently, we have
\begin{align}
\|{\rm\mathbf{B}}_n\tilde{\bm{\theta}}\|_2=1, n=1,2,\cdots,N,
\end{align}
where ${\rm\mathbf{B}}_n=[\mathbf{e}_n,\mathbf{e}_{n+N}]^T$, and $\mathbf{e}_n\in\mathbb{R}^{2N}$ denotes the $n$th column of $\mathbf{I}_{2N}$.

In this paper, we consider the joint beamforming design to minimize transmit power based on the assumption of imperfect CSI. By defining $\bar{\bm{{\rm H}}}_k\triangleq\mathcal{T}(\hat{\bm{{\rm H}}}_k)\in\mathbb{R}^{2N\times 2M}$ and $\bar{\bm{\Delta}}_k\triangleq\mathcal{T}(\bm{\Delta}_k)\in\mathbb{C}^{2N\times 2M}$, we have $\tilde{\bm{{\rm H}}}_k=\bar{\bm{{\rm H}}}_k+\bar{\bm{\Delta}}_k$. Thus, this problem can be mathematically formulated as
\begin{subequations}\label{A}
\begin{alignat}{2}
\min_{\tilde{{\rm\mathbf{x}}}, \tilde{\bm{\theta}}} \quad & \|\tilde{{\rm\mathbf{x}}}\|_2^2 \label{Aa}\\\
\mbox{s.t.}\quad
&\tilde{\bm{\theta}}^T\bm{{\rm D}}_{k,i}(\bar{\bm{{\rm H}}}_k+\bar{\bm{\Delta}}_k)\tilde{\mathbf{x}}\geq\xi_{k,i}, \forall \|\bar{\mathbf{\Delta}}_k\|_F\leq\delta_k,\forall k, i, \label{Ab}\\
&\|{\rm\mathbf{B}}_n\tilde{\bm{\theta}}\|_2=1, \forall n,\label{Ac}
\end{alignat}
\end{subequations}
where $\xi_{k,i}\triangleq\sigma_k\sqrt{\gamma_k}b_{k,i}$.

We can notice that the constraints over the phase shift vector in \eqref{Ac} are non-convex and difficult to be addressed. Additionally, the phase shift vector $\tilde{\bm{\theta}}$ couples with the transmit beamformer $\tilde{\mathbf{x}}$ in \eqref{Ab}, and the channel uncertainty is also taken into account. Therefore, problem \eqref{A} will be reformulated in the next subsection to obtain a simplified form.

\subsection{Problem Reformulation}
Let us focus on the constraints \eqref{Ab} which determine the CIR of the constellation point for each user. Due to the  criterion of worst-case robustness, the constraints \eqref{Ab} are equivalent to
\begin{align}
\min_{i=1,\cdots,S_k}\bigg[\tilde{\bm{\theta}}^T\bm{{\rm D}}_{k,i}\bar{\mathbf{\Delta}}_k{\rm\tilde{\mathbf{x}}}-(\xi_{k,i}-\tilde{\bm{\theta}}^T\bm{{\rm D}}_{k,i}\bar{\rm\mathbf{H}}_k{\rm\tilde{\mathbf{x}}})\bigg]\geq0,\forall \|\bar{\mathbf{\Delta}}_k\|_F\leq\delta_k, \forall k.
\end{align}
It should be noted that the CSI error matrix $\bar{\mathbf{\Delta}}_k$ is independent from each other. Thus, for the user $k$, the CIR constraints (17) can be translated into
\begin{align}
\min_{\forall i}\bigg[\left(\inf_{\|\bar{\mathbf{\Delta}}_k\|_F\leq\delta_k}\tilde{\bm{\theta}}^T\bm{{\rm D}}_{k,i}\bar{\mathbf{\Delta}}_k{\rm\tilde{\mathbf{x}}}\right)-(\xi_{k,i}-\tilde{\bm{\theta}}^T\bm{{\rm D}}_{k,i}\bar{\rm\mathbf{H}}_k{\rm\tilde{\mathbf{x}}})\bigg]\geq0.
\end{align}
For each $k$, it can be shown that \cite{8647896}
\begin{align}
\inf_{\|\bar{\mathbf{\Delta}}_k\|_F\leq\delta_k} \tilde{\bm{\theta}}^T\bm{{\rm D}}_{k,i}\bar{\mathbf{\Delta}}_k{\rm\tilde{\mathbf{x}}}=-\delta_k\|\tilde{\mathbf{x}}\|_2\|\tilde{\bm{\theta}}^T\bm{{\rm D}}_{k,i}\|_2, \forall i.
\end{align}
Thus, we can turn (18) to a compact form as
\begin{align}
\delta_k\|\tilde{\mathbf{x}}\|_2\|\tilde{\bm{\theta}}^T\bm{{\rm D}}_{k,i}\|_2\leq\tilde{\bm{\theta}}^T\bm{{\rm D}}_{k,i}\bar{\rm\mathbf{H}}_k{\rm\tilde{\mathbf{x}}}-\xi_{k,i}, \forall i.
\end{align}

By substituting (20) into problem \eqref{A}, we arrive at the problem as follows
\begin{subequations}\label{D}
\begin{align}
\min_{\tilde{{\mathbf{x}}}, \tilde{\bm{\theta}}} \quad & \|\tilde{{\rm\mathbf{x}}}\|_2^2\label{Da}\\
\mbox{s.t.}\quad
&\delta_k\|\tilde{\mathbf{x}}\|_2\|\tilde{\bm{\theta}}^T\bm{{\rm D}}_{k,i}\|_2\leq\tilde{\bm{\theta}}^T\bm{{\rm D}}_{k,i}\bar{\rm\mathbf{H}}_k{\rm\tilde{\mathbf{x}}}-\xi_{k,i}, \forall k, i.\label{Db}\\
&\|{\rm\mathbf{B}}_n\tilde{\bm{\theta}}\|_2=1, \forall n.\label{Dc}
\end{align}
\end{subequations}

Nevertheless, problem \eqref{D} is still not a convex optimization problem. It is shown that $\tilde{\mathbf{x}}$ and $\tilde{{\rm\bm{\theta}}}$ couple with each other in two different forms in the constraints \eqref{Db}. Furthermore, the constant-modulus constraints over the phase shift vector $\tilde{\bm{\theta}}$ are also difficult to tackle.

In what follows, the problem is first solved by considering a single-user case with binary phase-shift keying (BPSK) constellation, and then we will extend it to the general multiuser case.

\section{Single-user Case with BPSK}

In this section, we consider the case of single-user system with BPSK constellation. For the BPSK constellation, the CIR is established by only one hyperplane for each constellation point. So, we have $k=S_k=1$, and we drop the subscripts for simplicity. Thus, problem \eqref{D} reduces to
\begin{subequations}\label{S}
\begin{align}
\min_{\tilde{{\mathbf{x}}}, \tilde{\rm\bm{\theta}}} \quad & \|\tilde{\rm\mathbf{x}}\|_2\label{Sa}\\
\mbox{s.t.}\quad
&\delta\|\tilde{\mathbf{x}}\|_2\|\tilde{\bm{\theta}}^T\bm{{\rm D}}\|_2\leq\tilde{\bm{\theta}}^T\bm{{\rm D}}\bar{\rm\mathbf{H}}{\rm\tilde{\mathbf{x}}}-\xi, \label{Sb}\\
&\|{\rm\mathbf{B}}_n\tilde{\bm{\theta}}\|_2=1, \forall n.\label{Sc}
\end{align}
\end{subequations}

Since the BPSK constellation is adopted, based on (13), we have $\mathbf{D}=\text{diag}(2,-2)\otimes\mathbf{I}_N$. Besides, according to the definition of $\tilde{\bm{\theta}}$, we know $\|\tilde{\bm{\theta}}\|_2=\sqrt{N}$ and then we have $\|\tilde{\bm{\theta}}^T\bm{{\rm D}}\|_2=2\sqrt{N}$. Therefore, problem \eqref{S} can be transformed into
\begin{subequations}\label{S1}
\begin{align}
\min_{\tilde{{\mathbf{x}}}, \tilde{\rm\bm{\theta}}} \quad & \|\tilde{\rm\mathbf{x}}\|_2\label{S1a}\\
\mbox{s.t.}\quad
&2\delta\sqrt{N}\|\tilde{\mathbf{x}}\|_2\leq\tilde{\bm{\theta}}^T\bm{{\rm D}}\bar{\rm\mathbf{H}}{\rm\tilde{\mathbf{x}}}-\xi, \label{S1b}\\
&\|{\rm\mathbf{B}}_n\tilde{\bm{\theta}}\|_2=1, \forall n.\label{S1c}
\end{align}
\end{subequations}

In order to solve this problem, we propose two approximate algorithms, namely the SDR-based method and the AO-based method, respectively. As can be seen later, both algorithms can provide valuable insights for the multiuser case.

\subsection{SDR-based Method}

By analyzing problem \eqref{S1}, we know that the optimal beamforming scheme for any $\tilde{\rm\bm{\theta}}$ is the maximum ratio transmission (MRT). Let $P\triangleq\|\tilde{\rm\mathbf{x}}\|_2>0$, and thus the optimal beamformer for the BS can be expressed as $\tilde{\rm\mathbf{x}}^*=P\tilde{\bm{\theta}}^T\bm{{\rm D}}\bar{\rm\mathbf{H}}/\|\tilde{\bm{\theta}}^T\bm{{\rm D}}\bar{\rm\mathbf{H}}\|_2$. Based on this fact, problem \eqref{S1} can be rewritten as
\begin{subequations}\label{S2}
\begin{align}
\min_{P, \tilde{{\bm{\theta}}}} \quad & P\label{S2a}\\
\mbox{s.t.}\quad
&\frac{\xi}{P}+2\delta\sqrt{N}\leq\|\tilde{\bm{\theta}}^T\bm{{\rm D}}\bar{\rm\mathbf{H}}\|_2, \label{S2b}\\
&\|{\rm\mathbf{B}}_n\tilde{\bm{\theta}}\|_2=1, \forall n.\label{S2c}\\
&P>0\label{S2d}.
\end{align}
\end{subequations}

Although the optimal structure of the transmit beamformer $\tilde{\mathbf{x}}$ is given, it is still not straightforward to solve problem \eqref{S2} due to the non-convex constraints \eqref{S2b} and \eqref{S2c}. However, it can be readily proved by the strict monotonicity that the constraint \eqref{S2b} holds with equality at the optimal solution. Thus, the optimal transmit power can be given by
\begin{align}
P^*=\frac{\xi}{\|\tilde{\bm{\theta}}^T\bm{{\rm D}}\bar{\rm\mathbf{H}}\|_2-2\delta\sqrt{N}}
\end{align}
for any feasible $\tilde{\bm{\theta}}$. Then, the power minimization problem \eqref{S2} is equivalent to
\begin{subequations}\label{S3}
\begin{align}
\max_{\tilde{{\bm{\theta}}}} \quad & \|\tilde{\bm{\theta}}^T\bm{{\rm D}}\bar{\rm\mathbf{H}}\|_2^2\label{S3a}\\
\mbox{s.t.}\quad
&\|{\rm\mathbf{B}}_n\tilde{\bm{\theta}}\|_2=1,\forall n.\label{S3b}
\end{align}
\end{subequations}

Upon defining $\tilde{\bm{\Theta}}\triangleq\tilde{\bm{\theta}}\tilde{\bm{\theta}}^T$, we can further reformulate problem \eqref{S3} as
\begin{subequations}\label{S5}
\begin{align}
\max_{\tilde{{\bm{\Theta}}}} \quad & \text{tr}(\bm{{\rm D}}\bar{\rm\mathbf{H}}\bar{\rm\mathbf{H}}^T\bm{{\rm D}}^T\tilde{\bm{\Theta}})\label{S5a}\\
\mbox{s.t.}\quad
&\text{tr}({\rm\mathbf{B}}_n^T{\rm\mathbf{B}}_n\tilde{\bm{\Theta}})=1, \forall n.\label{S5b}\\
&\mathbf{\tilde{\Theta}}\succeq0\label{S5c},\\
&\text{rank}(\tilde{\mathbf{\Theta}})=1.\label{S5d}
\end{align}
\end{subequations}
By leveraging on the technique of SDR, i.e., dropping the rank-one constraint \eqref{S5d}, we can approximately solve problem \eqref{S5} by solving the following semidefinite programming (SDP) problem
\begin{subequations}\label{S6}
\begin{align}
\max_{\tilde{{\bm{\Theta}}}} \quad & \text{tr}(\bm{{\rm D}}\bar{\rm\mathbf{H}}\bar{\rm\mathbf{H}}^T\bm{{\rm D}}^T\tilde{\bm{\Theta}})\label{S6a}\\
\mbox{s.t.}\quad
&\text{tr}({\rm\mathbf{B}}_n^T{\rm\mathbf{B}}_n\tilde{\bm{\Theta}})=1,\forall n,\label{S6b}\\
&\mathbf{\tilde{\Theta}}\succeq0\label{S6c},
\end{align}
\end{subequations}
which is convex and can be solved efficiently by, e.g., interior point method.

\begin{Lemma}
Problem \eqref{S6} admits an infinite number of optimal solutions. Besides, if $\tilde{\bm{\Theta}}^*$ is an optimal solution to problem \eqref{S6} with two eigenvectors being
\begin{align}
\tilde{\bm{\theta}}_1^*\triangleq[{\tilde{\bm{\theta}}_{11}}^T,~{\tilde{\bm{\theta}}_{12}}^T]^T\in\mathbb{R}^{2N}
\end{align}
and
\begin{align}
\tilde{\bm{\theta}}_2^*\triangleq[-{\tilde{\bm{\theta}}_{12}}^T,~{\tilde{\bm{\theta}}_{11}}^T]^T\in\mathbb{R}^{2N},
\end{align}
then a new optimal solution can be constructed with a lower rank than that of $\tilde{\bm{\Theta}}^*$.
\end{Lemma}

\begin{IEEEproof}
See Appendix A.
\end{IEEEproof}

In general, the optimal solution is not rank-one. When an optimal solution $\tilde{\bm{\Theta}}^*$ to problem \eqref{S6} is obtained, we can perform the rank reduction operation based on Lemma 1 first. Then the Gaussian randomization procedure \cite{5447068} can be applied to generate an approximate solution if the rank is still not rank-one.

\subsection{AO-Based Method}

To obtain a unique rank-1 solution with low complexity, we propose an AO-based method to solve problem \eqref{S1}. To begin with, let $\bar{\mathbf{x}}\triangleq\tilde{\mathbf{x}}/P$, we can convert problem \eqref{S1} into
\begin{subequations}\label{S7}
\begin{align}
\min_{P, \bar{\mathbf{x}},\tilde{\rm\bm{\theta}}} \quad & P\label{S7a}\\
\mbox{s.t.}\quad
&\frac{\xi}{P}+2\delta\sqrt{N}\leq\tilde{\bm{\theta}}^T\bm{{\rm D}}\bar{\rm\mathbf{H}}{\rm\bar{\mathbf{x}}}, \label{S7b}\\
&\|{\rm\mathbf{B}}_n\tilde{\bm{\theta}}\|_2=1,\forall n,\label{S7c}\\
&\|\bar{\mathbf{x}}\|_2=1.\label{S7d}
\end{align}
\end{subequations}
Then, we will alternately optimize the transmit beamformer and the phase shift vector.

\subsubsection{Transmit Beamforming Design}
For any given $\tilde{\rm\bm{\theta}}$, the optimal transmit beamformer can be obtained by the MRT, and thus the minimum transmit power is given by
\begin{align}
P^*=\xi/(\|\tilde{\bm{\theta}}^T\bm{{\rm D}}\bar{\rm\mathbf{H}}\|-2\delta\sqrt{N})
\end{align}
such that \eqref{S7b} holds with equality. Meanwhile, the MRT transmit beamformer to problem \eqref{S1} can be expressed as
\begin{align}
\tilde{\mathbf{x}}^*=P^*\tilde{\bm{\theta}}^T\bm{{\rm D}}\bar{\rm\mathbf{H}}/\|\tilde{\bm{\theta}}^T\bm{{\rm D}}\bar{\rm\mathbf{H}}\|.
\end{align}

\subsubsection{Phase Shift Vector Design}
With a fixed $\tilde{\mathbf{x}}$, the problem of optimizing the phase shift vector can be transformed into
\begin{subequations}\label{S8}
\begin{align}
\max_{\tilde{\rm\bm{\theta}}} \quad & \tilde{\bm{\theta}}^T\bm{{\rm D}}\bar{\rm\mathbf{H}}{\rm\tilde{\mathbf{x}}}\label{S8a}\\
\mbox{s.t.}\quad
&\|{\rm\mathbf{B}}_n\tilde{\bm{\theta}}\|_2=1,\forall n.\label{S8b}
\end{align}
\end{subequations}
However, due to the constraints \eqref{S8b} which are typical kind of constant-envelope constraints, problem \eqref{S8} is difficult to solve.

To resolve this issue, let us revisit the objective in problem \eqref{S8}. Without loss of generality, we assume that the desired symbol is $s=1$, and thus we have $\mathbf{a}=[2,~0]^T$. Recall that
\begin{align}
&\!\!\!\!\tilde{\bm{\theta}}^T\bm{{\rm D}}\bar{\rm\mathbf{H}}{\rm\tilde{\mathbf{x}}}=\mathbf{a}^T\mathbf{\Theta}\bar{\Hb}{\rm\tilde{\mathbf{x}}}
=~\mathbf{a}^T\!
  \begin{bmatrix}
    \Re(\bm{\theta}^H) & \!\!-\Im(\bm{\theta}^H) \\
    \Im(\bm{\theta}^H) & \!\!\Re(\bm{\theta}^H) \\
  \end{bmatrix}\!\!
  \begin{bmatrix}
    \Re(\hat{\bm{{\rm H}}}) & \!\!-\Im(\hat{\bm{{\rm H}}}) \\
    \Im(\hat{\bm{{\rm H}}}) &\!\! \Re(\hat{\bm{{\rm H}}}) \\
  \end{bmatrix}\!\!
  \begin{bmatrix}
    \Re(\bm{{\rm x}}) \\
    \Im(\bm{{\rm x}}) \\
  \end{bmatrix}
=~2\Re(\bm{\theta}^H\hat{\bm{{\rm H}}}\bm{{\rm x}}).
\end{align}
Then, problem \eqref{S8} can be recast as
\begin{subequations}\label{S9}
\begin{align}
\max_{{\rm\bm{\theta}}} \quad & \Re(\bm{\theta}^H\hat{\bm{{\rm H}}}\bm{{\rm x}})\label{S9a}\\
\mbox{s.t.}\quad
&|{\rm\theta}_n|=1,\forall n.\label{S9b}
\end{align}
\end{subequations}
We can observe that the objective is the real part of the desired noise-free signal. It is always beneficial to improve the SER performance of the BPSK modulation scheme by maximizing the real part of the desired signal. For facilitating the algorithm development, we introduce the following lemma.

\begin{Lemma}
Problem \eqref{S9} is equivalent to
\begin{subequations}\label{S10}
\begin{align}
\max_{{\rm\bm{\theta}}} \quad & \bm{\theta}^H\hat{\bm{{\rm H}}}\bm{{\rm x}}\label{S10a}\\
\mbox{\rm {s.t.}}\quad
&|{\rm\theta}_n|=1,\forall n.\label{S10b}
\end{align}
\end{subequations}
\end{Lemma}

\begin{IEEEproof}
Assume that there is an optimal phase shift vector $\bm{\theta}^*$ to problem \eqref{S9} such that $\Im((\bm{\theta}^*)^H\hat{\bm{{\rm H}}}\bm{{\rm x}})\neq0$. Then, we can always perform a phase rotation on $\bm{\theta}^*$ and obtain a feasible solution $\bm{\theta}^{\star}=\bm{\theta}^*e^{\angle(\bm{\theta}^*)^H\hat{\bm{{\rm H}}}\bm{{\rm x}}}$. Moreover, we have
\begin{align}
\Re\left((\bm{\theta}^{\star})^H\hat{\bm{{\rm H}}}\bm{{\rm x}}\right)&=(\bm{\theta}^*)^H\hat{\bm{{\rm H}}}\bm{{\rm x}}e^{-\angle(\bm{\theta}^*)^H\hat{\bm{{\rm H}}}\bm{{\rm x}}}=|(\bm{\theta}^*)^H\hat{\bm{{\rm H}}}\bm{{\rm x}}|>\Re\left((\bm{\theta}^*)^H\hat{\bm{{\rm H}}}\bm{{\rm x}}\right).
\end{align}
It contradicts our assumption that $\bm{\theta}^*$ is the optimal solution to problem \eqref{S9}. Therefore, the optimal phase shift vector $\bm{\theta}^*$ to problem \eqref{S9} guarantees that $(\bm{\theta}^*)^H\hat{\bm{{\rm H}}}\bm{{\rm x}}$ is a real value. This completes the proof of Lemma 2.
\end{IEEEproof}

For problem \eqref{S10}, a closed-form solution can be obtained as follows
\begin{align}
\theta_n^*=e^{\angle[\hat{\bm{{\rm H}}}\bm{{\rm x}}]_n},\forall n,
\end{align}
where $[\hat{\bm{{\rm H}}}\bm{{\rm x}}]_n$ denotes the $n$th element of $\hat{\bm{{\rm H}}}\bm{{\rm x}}$. Then, we have $\tilde{\bm{\theta}}^*=[\Re(\bm{\theta}^*)^T,\Im(\bm{\theta}^*)^T]^T$ for the next iteration. Finally, the AO-based method is summarized in Algorithm 1.

\begin{algorithm}[t]
	\caption{: AO-based method for the single-user case}
	\label{alg:1}
	\begin{algorithmic}[1]
		\STATE initialize $\tilde{\rm{\bm{\theta}}}^0$, an accuracy threshold $\epsilon$, and set iteration index $t=0$.
        \REPEAT
        \STATE set $t:=t+1$.
		\STATE update the transmit power $P^t$ and transmit beamformer $\tilde{\bf{x}}^t$ with given $\tilde{\bm{\theta}}^{t-1}$ by (32), (33).
        \STATE update the phase shift vector $\bm{\theta}^t$ by (39) with given $\tilde{\mathbf{x}}^t$.
        \STATE update $\tilde{\bm{\theta}}^t=[\Re(\bm{\theta}^t)^T,\Im(\bm{\theta}^t)^T]^T$.
        \UNTIL if $(P^{t-1}-P^{t})/P^{t-1}<\epsilon$ is satisfied.
	\end{algorithmic}
\end{algorithm}

\subsection{Performance and Complexity Comparison}
In this subsection, we give a comparison between the proposed two methods in terms of the performance and the computational complexity.

In Fig. 3(a),  we can see that the average transmit power obtained by the SDR-based method and the AO-based method is almost the same for all $N$ in our simulations. Besides, we consider the lower bound of the transmit power which is obtained by solving problem \eqref{S6} without applying the Gaussian randomization procedure. It is shown that the AO-based method can achieve the near-optimal performance. In our simulations, all the optimal solutions obtained by the CVX \cite{cvx} to problem \eqref{S6} are rank-2 and have two eigenvectors in the forms of (29) and (30). Then, based on the Lemma 1, the rank-1 optimal solutions can be recovered from the rank-2 solutions. Thus, the global optimal solution to problem \eqref{S1} can be obtained by the SDR-based method in our simulations.

\begin{figure}[t]
\centering
\subfigure[]{
\begin{minipage}{7.5 cm}
\centering
\includegraphics[width=7.5 cm]{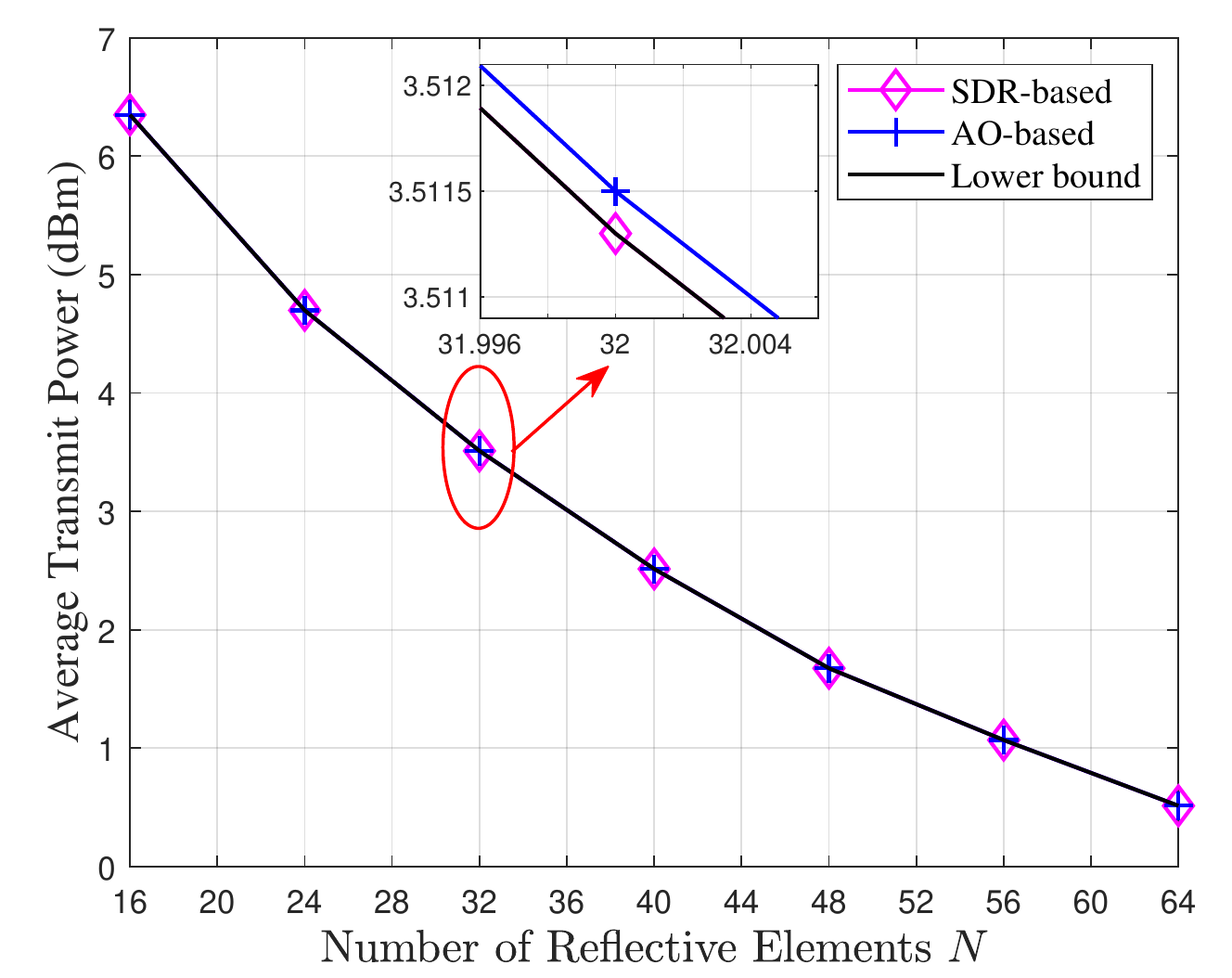}
\end{minipage}%
}
\subfigure[]{
\begin{minipage}{7.5 cm}
\centering
\includegraphics[width=7.5 cm]{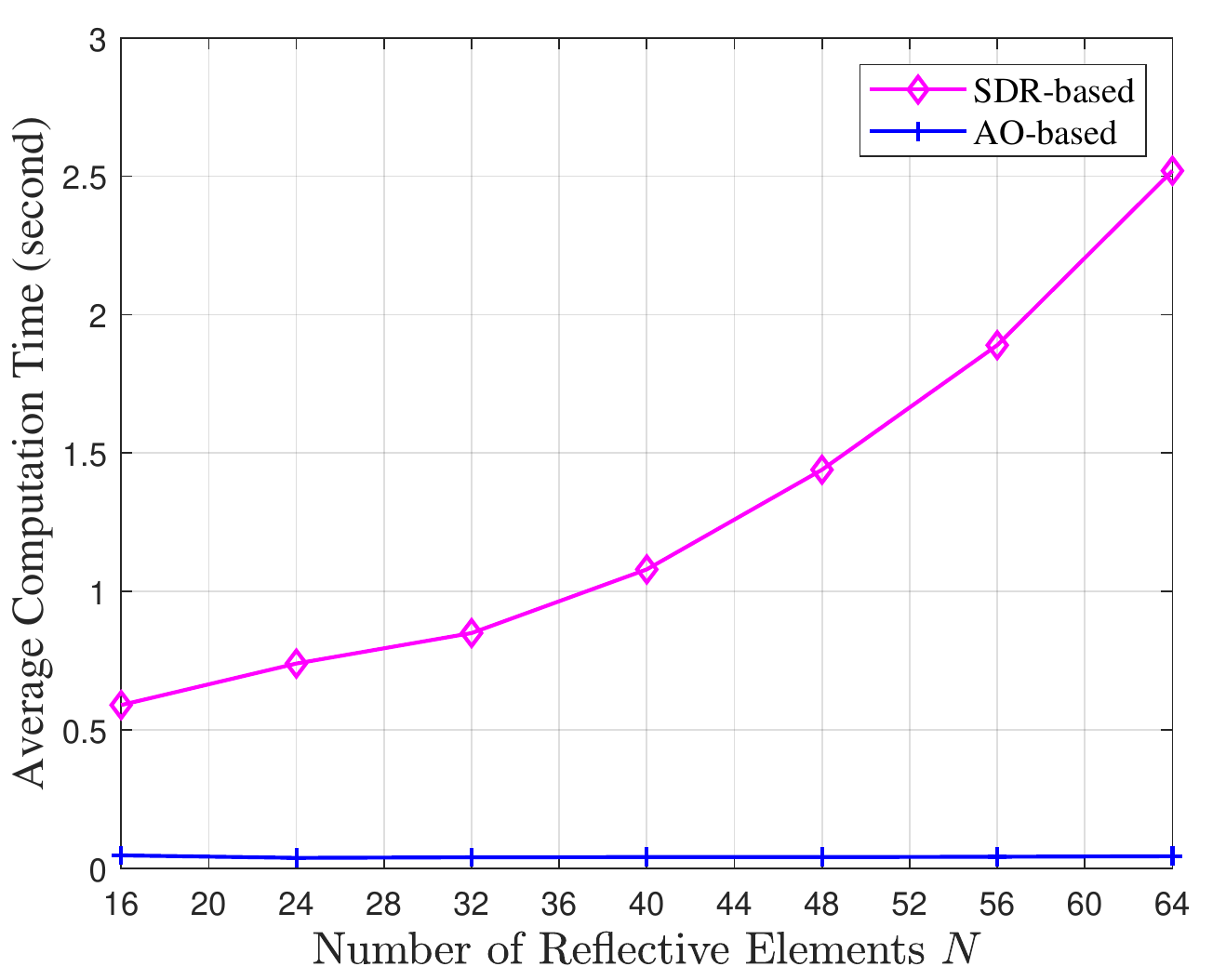}
\end{minipage}%
}
\caption{(a) Average transmit power versus the number of reflective elements $N$; (b) Average computation time versus the number of reflective elements $N$ ($M=4$, $\gamma=10$ dB, $\delta=0.02$).}
\vspace{-0.5cm}
\end{figure}

However, the two proposed methods achieve the similar performance at the cost of different computational complexity. Specifically, for the SDR-based method, the computational complexity comes from two parts. Firstly, the SDP problem \eqref{S6} can be solved with the complexity of $\mathcal{O}(N^3)$ \cite{souto2020exploiting}. Secondly, the extra Gaussian randomization procedure is performed with the complexity $\mathcal{O}(N^{3})$ \cite{6710599}. Hence, the total computational complexity of the SDR-based method is $\mathcal{O}(N^3)$. As a comparison, the complexity of the AO-based method is $\mathcal{O}(NM)$. Due to the fact that the number of reflective elements $N$ is greater than that of transmit antennas $M$ in general, the computational complexity of the SDR-based method is much larger than that of the AO-based method. As shown in Fig. 3(b), the computation time is almost constant as $N$ increases for the AO-based method. In contrast, we can observe that the average computation time increases remarkably with $N$ for the SDR-based method.

In summary, compared with the SDR-based method, a similar performance can be achieved by the AO-based method with a much lower computational complexity.

\section{Multiuser Case}
In this section, we investigate the multiuser scenario without direct links from the BS to the users, and then we also consider the case with the direct links. For solving problem \eqref{D} in the multiuser case, an algorithm based on the AO and the proximal gradient descent method \cite{beck2017first} is proposed.

\subsection{Beamforming Design without Direct Links}
For the multiuser scenario, one of the challenges for solving problem \eqref{D} lies in the constraints \eqref{Db}. The variables couple in two different forms, which brings a large difficulty to optimize $\tilde{\mathbf{x}}$ and $\tilde{\bm{\theta}}$ simultaneously. Inspired by the single-user case, we adopt the idea of the alternating optimization method to resolve this challenge. By optimizing the transmit beamformer and phase shift vector in an alternating manner, problem \eqref{D} can be split into two subproblems and solved iteratively.

\subsubsection{Transmit Beamforming Optimization}

For given $\tilde{\rm\bm{\theta}}^{t-1}$, where $t$ represents the iteration index, the subproblem of optimizing the transmit beamformer can be expressed as
\begin{subequations}\label{F2}
\begin{align}
\min_{\tilde{{\mathbf{x}}}} \quad & \|\tilde{\mathbf{x}}\|_2\label{F2a}\\
\mbox{s.t.}\quad
&(\tilde{\bm{\theta}}^{t-1})^T\bm{{\rm D}}_{k,i}\bar{\rm\mathbf{H}}_k{\rm\tilde{\mathbf{x}}}-\delta_k\|\tilde{\mathbf{x}}\|_2\|(\tilde{\bm{\theta}}^{t-1})^T\bm{{\rm D}}_{k,i}\|_2\geq\xi_{k,i}, \forall k,i.\label{F2b}
\end{align}
\end{subequations}
It can be verified that problem \eqref{F2} is a second-order cone program (SOCP) problem. Thus, problem \eqref{F2} can be optimally solved by some convex optimization solvers, such as CVX. For ease of description, we denote the optimal solution as $\tilde{\mathbf{x}}^t$ and let $P^t=\|\tilde{\mathbf{x}}^t\|_2$.

\subsubsection{Phase Shift Vector Optimization}
With a fixed $\tilde{\rm{\mathbf{x}}}^t$, problem \eqref{D} reduces to
\begin{subequations}\label{F3}
\begin{align}
\text{Find} \quad & \tilde{\rm{\bm{\theta}}}\label{F3a}\\
\mbox{s.t.}\quad
&\tilde{\bm{\theta}}^T\bm{{\rm D}}_{k,i}\bar{\rm\mathbf{H}}_k{\rm\tilde{\mathbf{x}}}^t-\delta_kP^t\|\tilde{\bm{\theta}}^T\bm{{\rm D}}_{k,i}\|_2\geq\xi_{k,i}, \forall k,i,\label{F3b}\\
&\|{\rm\mathbf{B}}_n\tilde{\bm{\theta}}\|_2=1, \forall n,\label{F3c}
\end{align}
\end{subequations}
which is a feasibility problem. Note that $\tilde{\bm{\theta}}^{t-1}$ is feasible to problem \eqref{F3}. Thus, to obtain an efficient update other than the trivial solution $\tilde{\bm{\theta}}^{t-1}$, an explicit objective should be constructed to guarantee the performance improvement for problem \eqref{F2} in terms of transmit power.

It can be proved that, at the optimal solution to problem \eqref{F2}, there is at least one constraint in \eqref{F2b} that holds with equality. Otherwise, we can always further reduce the transmit power. In view of this, a phase shift vector $\tilde{\bm{\theta}}$ which can enlarge the differences between the two sides of inequalities \eqref{F3b} is preferred. With the increase of the differences, the position of the received signal is pushed away from decision boundaries. It makes room for the reduction of the transmit power in the next iteration. As a result, we can update $\tilde{\rm\bm{\theta}}$ by solving the following problem
\begin{subequations}\label{F4}
\begin{align}
\max_{\tilde{\bm{\theta}}} \quad & f(\tilde{\bm{\theta}})\triangleq\sum_{k=1}^K\sum_{i=1}^{S_k}\left(\tilde{\bm{\theta}}^T\bm{{\rm D}}_{k,i}\bar{\rm\mathbf{H}}_k{\rm\tilde{\mathbf{x}}}^t-\delta_kP^t\|\tilde{\bm{\theta}}^T\bm{{\rm D}}_{k,i}\|_2\right)\label{F4a}\\
\mbox{s.t.}\quad
&\tilde{\bm{\theta}}^T\bm{{\rm D}}_{k,i}\bar{\rm\mathbf{H}}_k{\rm\tilde{\mathbf{x}}}^t-\delta_kP^t\|\tilde{\bm{\theta}}^T\bm{{\rm D}}_{k,i}\|_2\geq\xi_{k,i}, \forall k,i,\label{F4b}\\
&\|{\rm\mathbf{B}}_n\tilde{\bm{\theta}}\|_2^2=1, \forall n.\label{F4c}
\end{align}
\end{subequations}

Now, let us treat the non-convex constraints \eqref{F4c}. The main idea is relaxation and penalization. Specifically, we first relax the constant envelope constraints by
\begin{align}
\|{\bf{B}}_n\tilde{\bm{\theta}}\|_2^2\leq1, \forall n.
\end{align}
To restrict the gap caused by the relaxation of (43), we introduce a penalty term
\begin{align}
g(\tilde{\bm{\theta}})\triangleq\sum_{n=1}^N(\|\mathbf{B}_n\tilde{\bm{\theta}}\|_2^2-1)=\|\tilde{\bm{\theta}}\|_2^2-N
\end{align}
to the objective. Therefore, problem \eqref{F4} can be approximated by
\begin{subequations}\label{F5}
\begin{align}
\max_{\tilde{\bm{\theta}}} \quad & f(\tilde{\bm{\theta}})+\lambda g(\tilde{\bm{\theta}})\label{F5a}\\
\mbox{s.t.}\quad
&\tilde{\bm{\theta}}^T\bm{{\rm D}}_{k,i}\bar{\rm\mathbf{H}}_k{\rm\tilde{\mathbf{x}}}^t-\delta_kP^t\|\tilde{\bm{\theta}}^T\bm{{\rm D}}_{k,i}\|_2\geq\xi_{k,i}, \forall k,i,\label{F5b}\\
&\|{\rm\mathbf{B}}_n\tilde{\bm{\theta}}\|_2^2\leq1, \forall n,\label{F5c}
\end{align}
\end{subequations}
where $\lambda>0$ is a penalty factor that needs not be very large when $N$ is large.

For problem \eqref{F5}, let us focus on the non-concave objective due to the penalty term. Owing to the property of $g(\tilde{\bm{\theta}})$, we can solve the problem by applying the proximal gradient descent method and a stationary point can be obtained eventually \cite{beck2017first}. The details of the method are described as follows.

At the $t$th AO iteration, we introduce $\bm{\vartheta}_{\ell}$ as the approximate point of problem \eqref{F5} at the $\ell$th iteration by the PGD method. Then, the gradient of $g(\tilde{\bm{\theta}})$ at $\bm{\vartheta}_{\ell}$ is calculated as
\begin{align}
\nabla g(\bm{\vartheta}_{\ell})=2\bm{\vartheta}_{\ell}.
\end{align}
Next, we can update $\bm{\vartheta}$ by
\begin{align}\label{U2}
\bm{\vartheta}_{\ell+1}=\Pi_{\frac{1}{\beta}f,\tilde{\mathcal{V}}}(\bm{\vartheta}_{\ell}+\frac{\lambda}{\beta}\nabla g(\bm{\vartheta}_{\ell})),
\end{align}
where $\beta$ denotes the step size, $\tilde{\mathcal{V}}$ is the convex feasible set defined by \eqref{F5b}-\eqref{F5c}, and $\Pi_{\frac{1}{\beta}f,\tilde{\mathcal{V}}}(\cdot)$ is a proximal operator and defined as
\begin{align}
\Pi_{\frac{1}{\beta}f,\tilde{\mathcal{V}}}(\mathbf{x})=\arg\max_{\tilde{\bm{\theta}}\in\tilde{\mathcal{V}}}~\frac{1}{\beta}f(\tilde{\bm{\theta}})-\frac{1}{2}\|\tilde{\bm{\theta}}-\mathbf{x}\|^2_2.
\end{align}

Therefore, we can update $\bm{\vartheta}$ by solving the following convex optimization problem
\begin{subequations}\label{F6}
\begin{align}
\max_{\bm{\vartheta}} \quad & \frac{1}{\beta}f(\bm{\vartheta})-\frac{1}{2}\|\bm{\vartheta}-\mathbf{y}_{\ell+1}\|_2^2\label{F6a}\\
\mbox{s.t.}\quad
&\bm{\vartheta}^T\bm{{\rm D}}_{k,i}\bar{\rm\mathbf{H}}_k{\rm\tilde{\mathbf{x}}}^t-\delta_kP^t\|\bm{\vartheta}^T\bm{{\rm D}}_{k,i}\|_2\geq\xi_{k,i}, \forall k,i,\label{F6b}\\
&\|{\rm\mathbf{B}}_n\bm{\vartheta}\|_2^2\leq1, \forall n.\label{F6c}
\end{align}
\end{subequations}
In consequence, problem \eqref{F4} can be approximately solved by iteratively solving problem \eqref{F6}. By analyzing the smoothness of $g(\tilde{\bm{\theta}})$, the convergency of the PGD method can be guaranteed when $\beta>\lambda$, and the limit point of the sequence $\{\bm{\vartheta}_{\ell}\}$ is a stationary point of problem \eqref{F5} \cite{beck2017first}.

In summary, we can approximately solve problem \eqref{D} by solving subproblems \eqref{F2} and \eqref{F5} in an alternating manner. Then, to tackle the non-concave objective in problem \eqref{F5}, the PGD method is applied. The details of the proposed method are summarized in Algorithm 2, and its convergency can be guaranteed by the following proposition.

\begin{proposition}
The power sequence $\{P^t\}$ of problem \eqref{F2} obtained by Algorithm 2 is non-increasing.
\end{proposition}

\begin{IEEEproof}
For ease of exposition, we rewrite problem \eqref{F2} first. Upon defining
\begin{align}
h(\tilde{\mathbf{x}}|\tilde{\bm{\theta}})\triangleq\|\tilde{\mathbf{x}}\|_2+\delta_{\tilde{\mathcal{X}}}(\tilde{\mathbf{x}}),
\end{align}
where $\delta_{\tilde{\mathcal{X}}}(\cdot)$ is an indicator function, and $\tilde{\mathcal{X}}$ is the convex feasible set of $\tilde{\mathbf{x}}$ defined by \eqref{F2b} with given $\tilde{\bm{\theta}}$. Thus, problem \eqref{F2} amounts to the problem $\min_{\tilde{\mathbf{x}}}h(\tilde{\mathbf{x}}|\tilde{\bm{\theta}})$. Then, we have $P^t=\min_{\tilde{\mathbf{x}}}h(\tilde{\mathbf{x}}|\tilde{\bm{\theta}}^{t-1})=h(\tilde{\mathbf{x}}^t|\tilde{\bm{\theta}}^{t-1})$. Due to the constraints \eqref{F6b}, $\{\tilde{\bm{\theta}}^t, \tilde{\mathbf{x}}^t\}$ is always a feasible solution to problem \eqref{F2}, where $\tilde{\bm{\theta}}^t$ is obtained by steps 7-10 in Algorithm 2. Therefore, we have
\begin{align}
P^t=h(\tilde{\mathbf{x}}^t|\tilde{\bm{\theta}}^{t-1})=h(\tilde{\mathbf{x}}^t|\tilde{\bm{\theta}}^t)\geq\min_{\tilde{\mathbf{x}}}h(\tilde{\mathbf{x}}|\tilde{\bm{\theta}}^t)=P^{t+1}.
\end{align}
Thus, the power sequence $\{P^t\}$ is non-increasing. Furthermore, since the power sequence is lower bounded by $0$, the convergency of Algorithm 2 can be guaranteed. This completes the proof.
\end{IEEEproof}

\begin{algorithm}[t]
	\caption{: AO-based method for the multiuser case}
	\label{alg:1}
	\begin{algorithmic}[1]
		\STATE initialize $\tilde{\rm{\bm{\theta}}}^0$, accuracy tolerance $\epsilon_1$, $\epsilon_2$, and set the outer iteration index $t=0$.
        \REPEAT
        \STATE set $t:=t+1$.
		\STATE update $\tilde{\rm{\mathbf{x}}}^t$ and $P^t$ by solving problem \eqref{F2} with $\tilde{\rm{\bm{\theta}}}=\tilde{\rm{\bm{\theta}}}^{t-1}$.
        \STATE set the inner iteration index $\ell=0$.
        \STATE set $\bm{\vartheta}_{\ell}=\tilde{\bm{\theta}}^{t-1}$.
        \REPEAT
		\STATE set $\ell:=\ell+1$.
        \STATE update $\bm{\vartheta}$ by solving problem \eqref{F6} with given $\tilde{\rm{\mathbf{x}}}^t$ and $\bm{\vartheta}_{\ell-1}$.
        \UNTIL the decrease of the target value is less than $\epsilon_1$ and obtain the solution $\tilde{\bm{\theta}}^t=\bm{\vartheta}_{\ell}$.
        \UNTIL $(P^{t-1}-P^t)/P^{t-1}<\epsilon_2$.
	\end{algorithmic}
\end{algorithm}

\begin{remark}
In practice, due to the limitation of hardware implementation, the IRS elements can only support finite phase shift levels, e.g., $2^B$ phase shifts, where $B$ indicates the resolution of the IRS. Without loss of generality, we assume that the phase shifts are quantized uniformly and denote the set of discrete phase shifts as $\mathcal{C}\triangleq\{e^{j(\frac{2\pi}{2^B}m+\frac{\pi}{2^B}})|m=0,\cdots,2^B-1\}$. To address this challenge posed by the finite resolution, we adopt the method proposed in \cite{8811616} by replacing the set of discrete phase shifts with its convex hull. Hence, in the case of the finite resolution, problem \eqref{F4} can be approximated by
\begin{subequations}\label{F7}
\begin{align}
\max_{\tilde{\bm{\theta}}} \quad & f(\tilde{\bm{\theta}})+\lambda g(\tilde{\bm{\theta}})\label{F7a}\\
\mbox{s.t.}\quad
&\delta_kP^t\|\tilde{\bm{\theta}}^T\bm{{\rm D}}_{k,i}\|_2\leq\tilde{\bm{\theta}}^T\bm{{\rm D}}_{k,i}\bar{\rm\mathbf{H}}_k{\rm\tilde{\mathbf{x}}}^t-\xi_{k,i}, \forall k,i,\label{F7b}\\
&\tilde{\theta}_n+j\tilde{\theta}_{n+N}\in\bm{{\rm conv}}\mathcal{C},\forall n,\label{F7c}
\end{align}
\end{subequations}
where $\bm{{\rm conv}}\mathcal{C}$ denotes the convex hull of $\mathcal{C}$. Obviously, the constraints \eqref{F7c} are convex. Thus, the Algorithm 2 can be readily extended to the case of the finite resolution.
\end{remark}

\subsection{Beamforming Design with Direct Links}

Now, let us consider the scenario where the direct links from the BS to the users exist. In this case, the users can receive the signal not only reflected by the IRS but also from the BS. Thus, the signal received at the user $k$ can be written as
\begin{align}
y_k=\left(\bm{\theta}^H\text{diag}(\mathbf{h}_{{\rm r}k}^H)\bm{{\rm G}}+\mathbf{h}_{{\rm d}k}^H\right){\rm\mathbf{x}}+z_k,
\end{align}
where $\mathbf{h}_{{\rm r}k}\in\mathbb{C}^{N}$ is the channel from the IRS to the user $k$, and $\mathbf{h}_{{\rm d}k}\in\mathbb{C}^{M}$ denotes the direct link from the BS to the user $k$. For notational convenience, we define $\bm{{\rm H}}_{{\rm r}k}\triangleq\text{diag}(\mathbf{h}_{{\rm r}k}^H)\bm{{\rm G}}\in\mathbb{C}^{N\times M}$. Furthermore, we assume that both the reflective and the direct links can not be estimated accurately, and the channel estimations at the BS are given by
\begin{align}
\hat{\bm{{\rm H}}}_{{\rm r}k}=&~\bm{{\rm H}}_{{\rm r}k}-\bm{\Delta}_{{\rm r}k},\forall k\\
\hat{\mathbf{h}}_{{\rm d}k}=&~\mathbf{h}_{{\rm d}k}-\bm{\delta}_{{\rm d}k},\forall k,
\end{align}
where $\bm{\Delta}_{{\rm r}k}\in\mathbb{C}^{N\times M}$ and $\bm{\delta}_{{\rm d}k}\in\mathbb{C}^{M}$ are the estimation errors of the reflective link and the direct link, respectively. For the estimation errors, we also assume that they are Frobenius-norm bounded and therefore we have
\begin{align}
\|\bm{\Delta}_{{\rm r}k}\|_F\leq\frac{\sqrt{2}}{2}\delta_{{\rm r}k},\forall k\\
\|\bm{\delta}_{{\rm d}k}\|_2\leq\frac{\sqrt{2}}{2}\delta_{{\rm d}k},\forall k,
\end{align}
where $\delta_{{\rm r}k}$ and $\delta_{{\rm d}k}$ represent the radiuses of the estimation errors of the reflective and the direct links, respectively.

According to the same philosophy as in section $\rm\uppercase\expandafter{\romannumeral 2}$, the worst-case robust power minimization problem can be written as
\begin{subequations}\label{G}
\begin{align}
\min_{\tilde{{\rm\mathbf{x}}}, {\bm{\theta}}} \quad & \|\tilde{{\xb}}\|_2^2\label{Ga}\\
\mbox{s.t.}\quad
&{\rm \mathbf{A}}_k(\mathbf{\Theta}\tilde{\bm{{\rm H}}}_{{\rm r}k}+\tilde{\bm{{\rm H}}}_{{\rm d}k}){\rm\tilde{\mathbf{x}}}\succeq\bm{\xi}_k,~\forall k,\label{Gb}\\
&{\bf{\Theta}}=\mathcal{T}(\bm{\theta}^H),\label{Gc}\\
&|\theta_n|=1, \forall n,\label{Gd}
\end{align}
\end{subequations}
where $\tilde{\bm{{\rm H}}}_{{\rm r}k}=\mathcal{T}(\bm{{\rm H}}_{{\rm r}k})$ and $\tilde{\bm{{\rm H}}}_{{\rm d}k}=\mathcal{T}(\mathbf{h}_{{\rm d}k}^H)$. Then, substituting (54) and (55) into the constraints \eqref{Gb}, we have
\begin{align}
{\rm \mathbf{A}}_k(\mathbf{\Theta}\bar{\bm{\Delta}}_{{\rm r}k}+\bar{\bm{\Delta}}_{{\rm d}k}){\rm\tilde{\mathbf{x}}}\succeq\bm{\xi}_k-{\rm \mathbf{A}}_k(\mathbf{\Theta}\bar{\bm{{\rm H}}}_{{\rm r}k}+\bar{\bm{{\rm H}}}_{{\rm d}k}){\rm\tilde{\mathbf{x}}}, \forall \|\bar{\bm{\Delta}}_{{\rm r}k}\|_F\leq\delta_{{\rm r}k}, \|\bar{\bm{\Delta}}_{{\rm d}k}\|_F\leq\delta_{{\rm d}k},\forall k,
\end{align}
where $\bar{\bm{\Delta}}_{{\rm r}k}=\mathcal{T}(\bm{\Delta}_{{\rm r}k})$, $\bar{\bm{\Delta}}_{{\rm d}k}=\mathcal{T}(\bm{\delta}_{{\rm d}k})$, $\bar{\bm{{\rm H}}}_{{\rm r}k}=\mathcal{T}(\hat{\bm{{\rm H}}}_{{\rm r}k})$, and $\bar{\bm{{\rm H}}}_{{\rm d}k}=\mathcal{T}(\hat{\mathbf{h}}_{{\rm d}k})$. Following the same spirit as in Section $\rm\uppercase\expandafter{\romannumeral 2}$-D, the constraints (59) can be converted into
\begin{align}
&\|\tilde{\mathbf{x}}\|_2\left(\delta_{{\rm r}k}\|\tilde{\bm{\theta}}^T\bm{{\rm D}}_{k,i}\|_2+\delta_{{\rm d}k}\|\mathbf{a}_{k,i}\|_2\right)\leq(\tilde{\bm{\theta}}^T\bm{{\rm D}}_{k,i}\bar{\bm{{\rm H}}}_{{\rm r}k}+\mathbf{a}_{k,i}^T\bar{\bm{{\rm H}}}_{{\rm d}k}){\rm\tilde{\mathbf{x}}}-\xi_{k,i}, \forall k,i.
\end{align}
Therefore, problem \eqref{G} is equivalent to
\begin{subequations}\label{G1}
\begin{align}
\min_{\tilde{{\rm\mathbf{x}}}, \tilde{\bm{\theta}}} \quad & \|\tilde{{\rm\mathbf{x}}}\|_2^2\label{G1a}\\
\mbox{s.t.}\quad
&\|\tilde{\mathbf{x}}\|_2\left(\delta_{{\rm r}k}\|\tilde{\bm{\theta}}^T\bm{{\rm D}}_{k,i}\|_2+\delta_{{\rm d}k}\|\mathbf{a}_{k,i}\|_2\right)\leq(\tilde{\bm{\theta}}^T\bm{{\rm D}}_{k,i}\bar{\bm{{\rm H}}}_{{\rm r}k}+\mathbf{a}_{k,i}^T\bar{\bm{{\rm H}}}_{{\rm d}k}){\rm\tilde{\mathbf{x}}}-\xi_{k,i}, \forall k,i,\label{G1b}\\
&\|{\rm\mathbf{B}}_n\tilde{\bm{\theta}}\|_2=1, \forall n.\label{G1c}
\end{align}
\end{subequations}

Since an extended version of Algorithm 2 can be developed readily to solve this problem, we omit the details here.

\section{NUMERICAL RESULTS}

In this section, some numerical results are provided to evaluate the performance of the proposed algorithms.

\begin{figure}
\begin{center}
\includegraphics[width=7.5 cm]{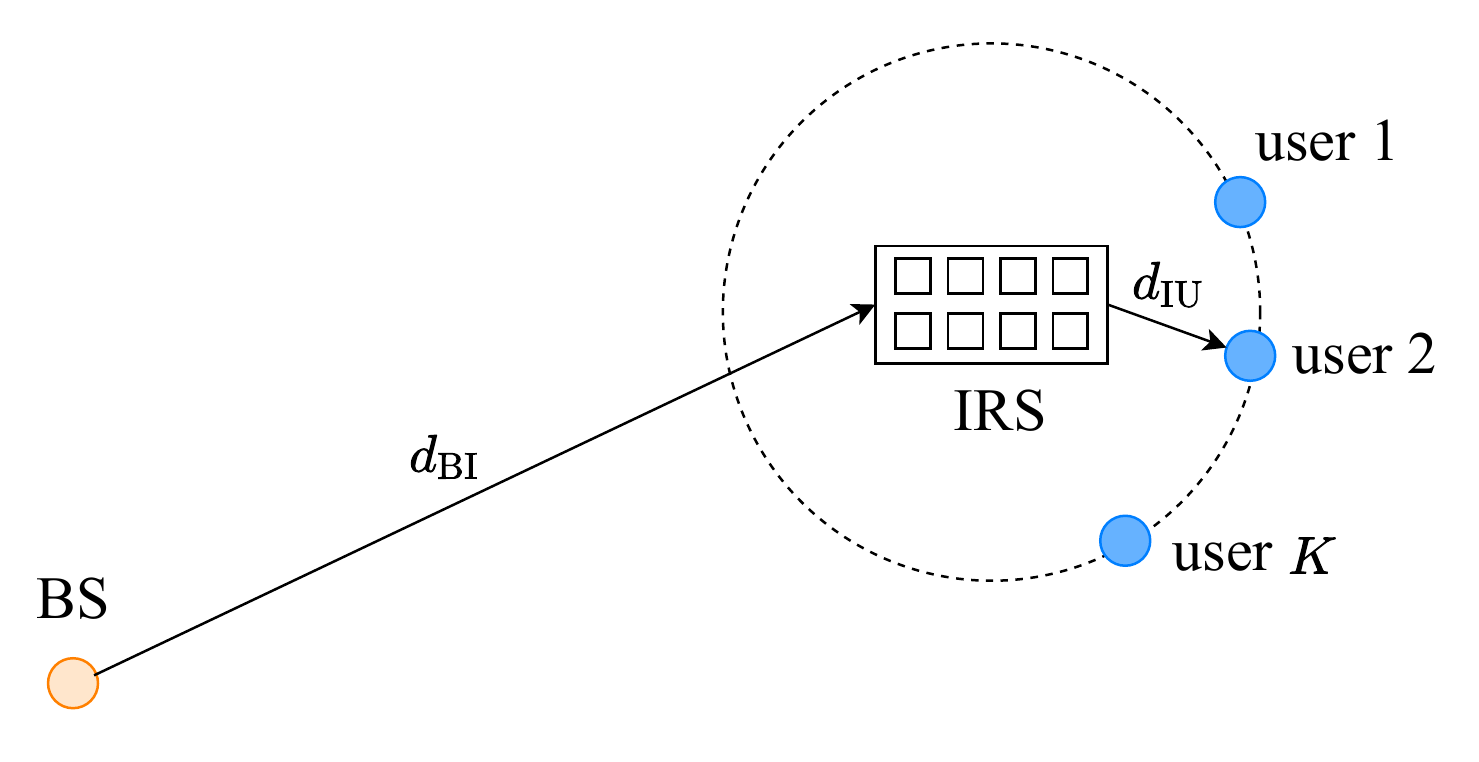}
\caption{Simulation setting for the IRS-aided communication system with $K$ users.}
\label{Casimir}
\end{center}
\vspace{-0.5cm}
\end{figure}

The simulation scenario is shown in Fig. 4, where the BS equipped with $M=4$ antennas is located at the origin point, while $K$ single-antenna users randomly locate on the dashed circle around the IRS panel. The IRS panel is expected to be deployed close to the users. Thus, the distances between the IRS and the users are set as $d_{\rm IU}=3$ m, and the distance from the BS to the IRS is $d_{\rm BI}=50$ m. We consider Rician fading for the channel $\mathbf{G}$ due to the existence of the LoS link from the BS to the IRS and Rayleigh fading for the channel $\mathbf{h}_k$ between the IRS and the user $k$. So, the BS-IRS channel can be expressed as

\begin{align}
\mathbf{G}=\sqrt{\frac{K_{\rm R}}{K_{\rm R}+1}}\mathbf{G}^{{\rm LoS}}+\sqrt{\frac{1}{K_{\rm R}+1}}\mathbf{G}^{{\rm NLoS}},
\end{align}
where $K_{\rm R}=3$ dB denotes the Rician factor, $\mathbf{G}^{{\rm LoS}}\in\mathbb{C}^{N\times M}$ is the deterministic component determined by the geometric settings, and $\mathbf{G}^{{\rm NLoS}}\in\mathbb{C}^{N\times M}$ is the component following the standard Rayleigh distribution. As for the large-scale fading, we consider a distance-dependent path loss model which is given by
\begin{align}
P_{\rm L}=C_0\left(\frac{d}{d_0}\right)^{-\alpha},
\end{align}
where $C_0=30$ dB represents the path loss at the reference distance $d_0=1$ m, $\alpha$ denotes the path loss exponent determined by the environment between the links, and $d$ is the distance between the transmitter and the receiver \cite{8811733}. We assume $\alpha_{\rm BI}=2.5$ for the BS-IRS link and $\alpha_{\rm IU}=2.8$ for the links between the IRS and the users. As for the direct links, we set $\alpha_{\rm BU}=3.5$ due to the rich scattering environment. Additionally, we set the noise power $\sigma^2_k=-80$ dBm for all $k$.

\begin{figure}
\begin{center}
\includegraphics[width=7.5 cm]{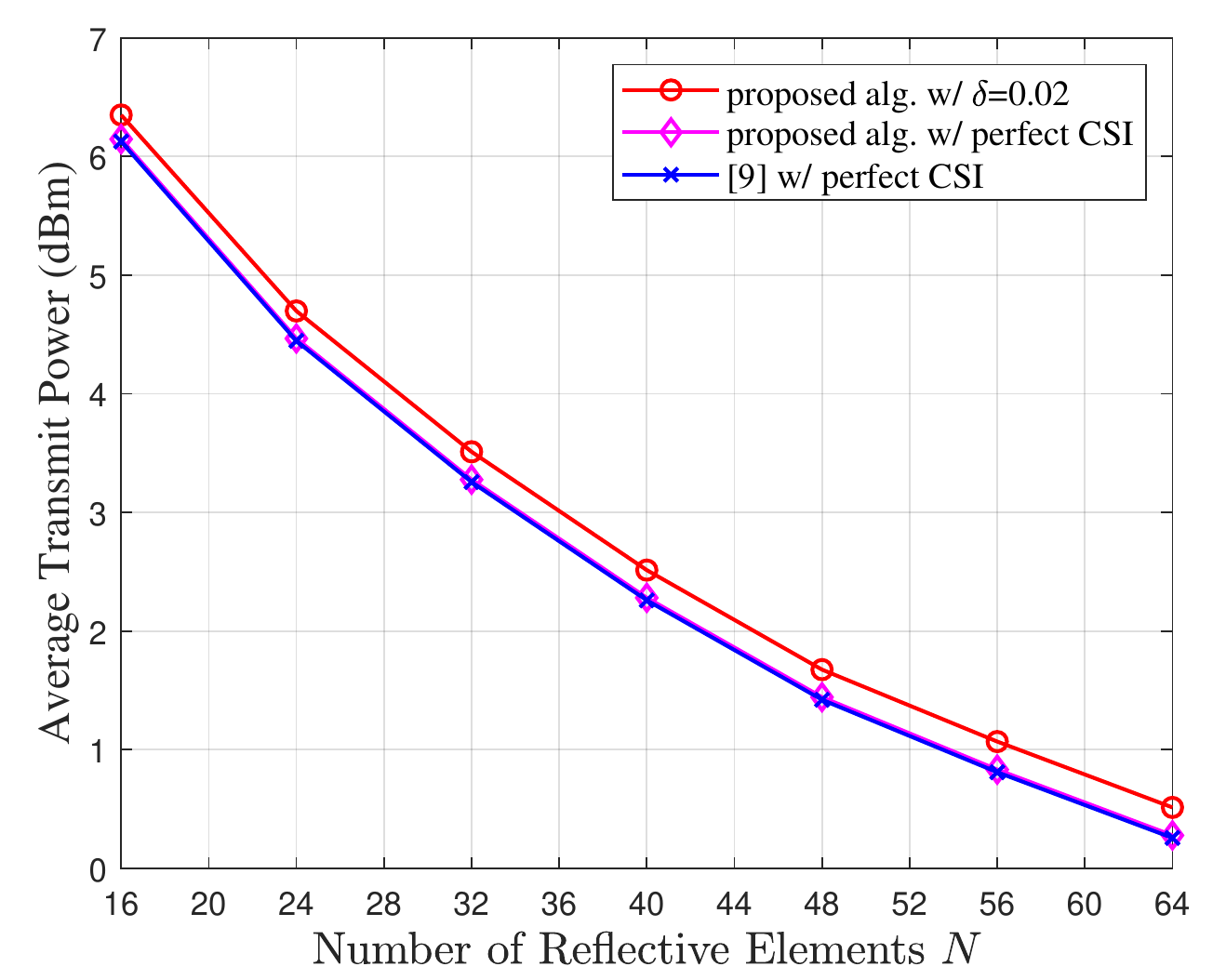}
\caption{Average transmit power versus the number of reflective elements $N$ ($M=4$, $\gamma=10$ dB).}
\label{Casimir}
\end{center}
\vspace{-0.5cm}
\end{figure}

\subsection{Single-user Case}

First, we consider the single-user case with the infinite-resolution IRS. From Fig. 5, we see that the average power decreases remarkably with the number of IRS elements $N$. As a comparison, we also consider the conventional BLP scheme which is studied in \cite{8811733}. As expected, we can observe that the average transmit power is the same for the BLP and SLP with a given $N$ when the BS has perfect CSI. The underlying cause is that the multiuser interference does not exist in a single-user system. Thus, the SLP can not provide an additional performance gain by converting MUI into the constructive energy compared with the conventional BLP. Besides, as shown in Fig. 5, the robustness of the network can be provided at the cost of extra power consumption when the accurate CSI is not available at the BS.

\begin{figure*}
\centering
\subfigure[QPSK]{
\begin{minipage}{0.3\linewidth}
\centering
\includegraphics[width=5.3 cm]{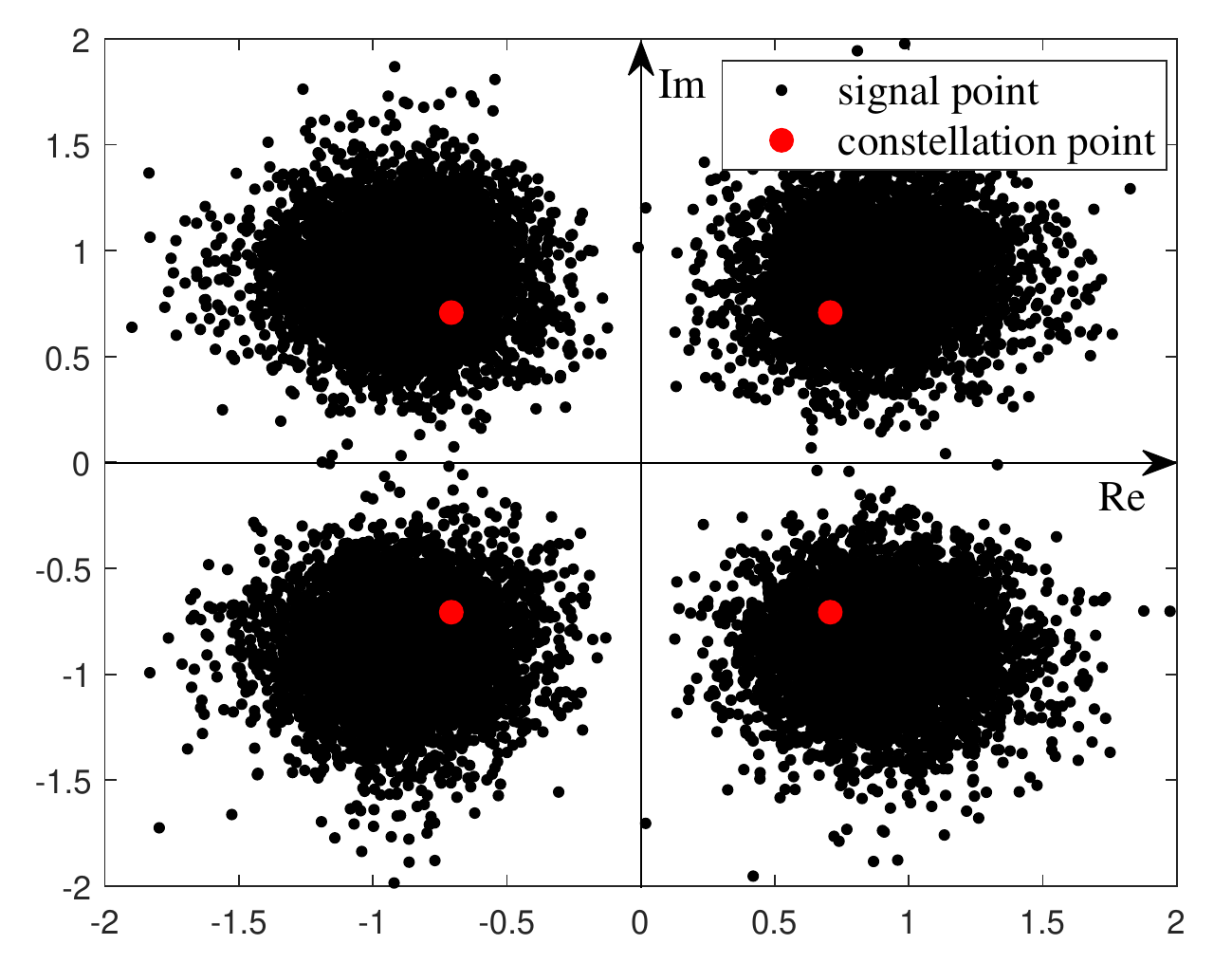}
\end{minipage}%
}
\subfigure[8-PSK]{
\begin{minipage}{0.3\linewidth}
\centering
\includegraphics[width=5.3 cm]{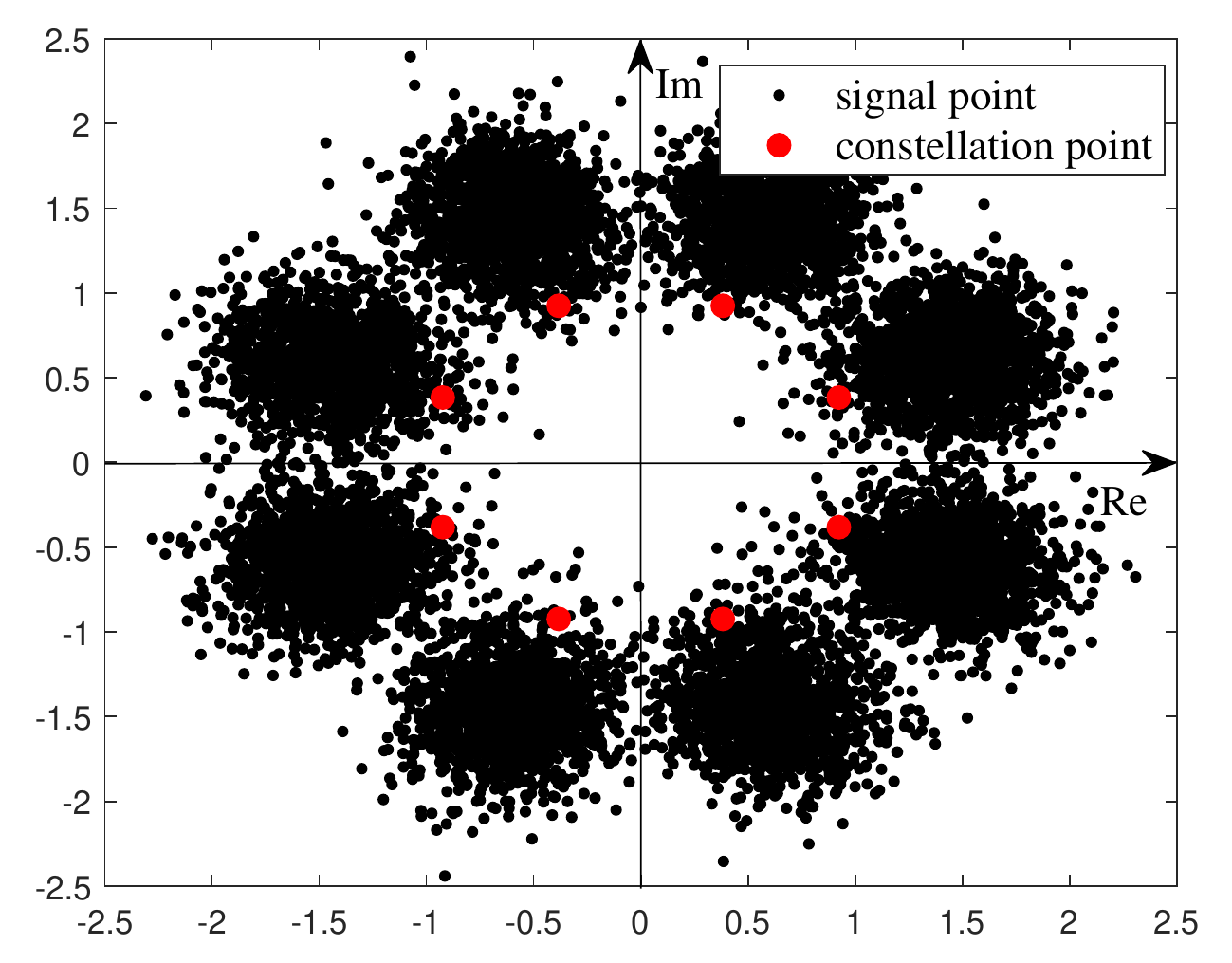}
\end{minipage}%
}
\subfigure[16-QAM]{
\begin{minipage}{0.3\linewidth}
\centering
\includegraphics[width=5.3 cm]{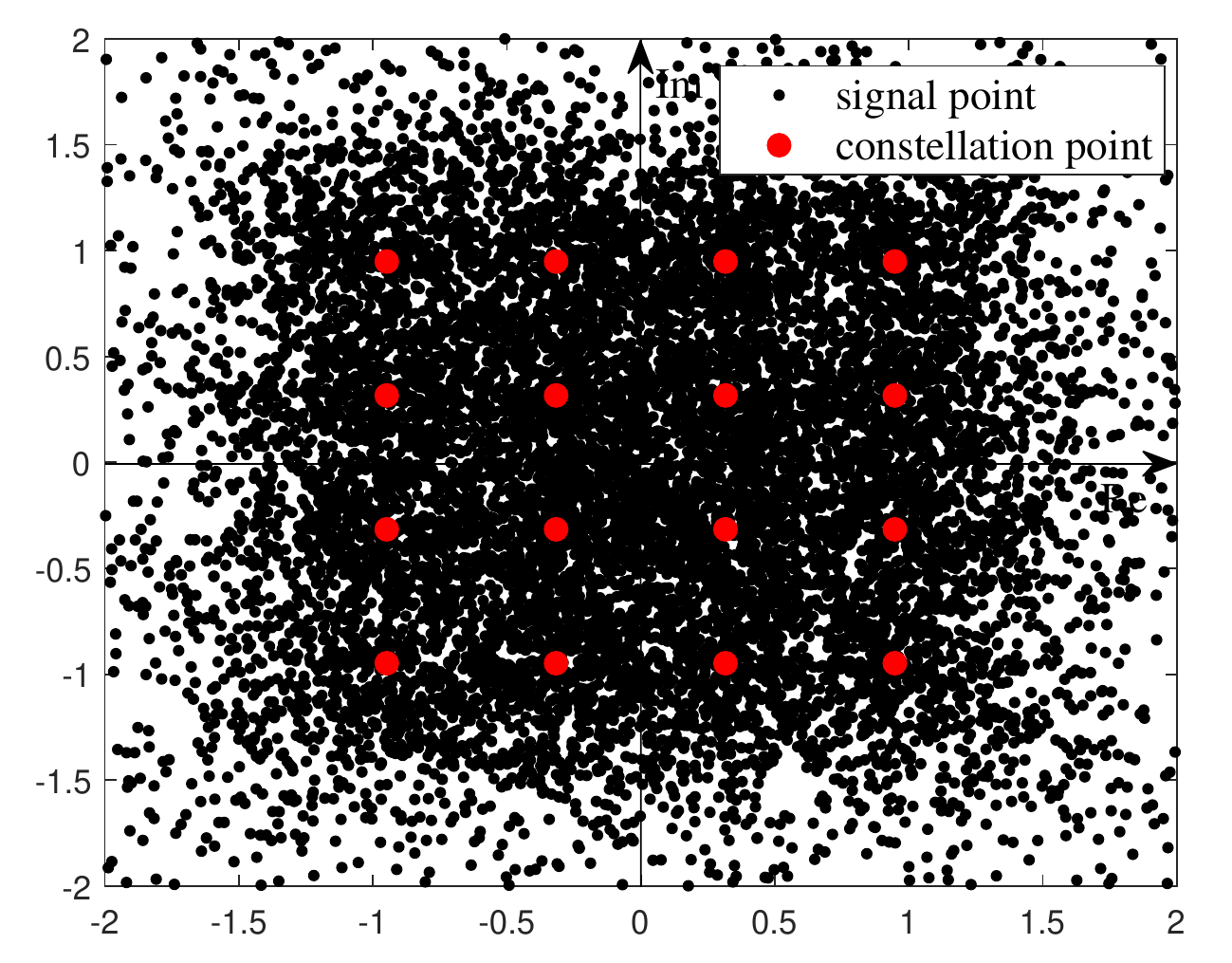}
\end{minipage}
}
\caption{The positions of the received signal at the users with different constellations ($N=64, B=\infty, \gamma=10$dB).}
\vspace{-0.5cm}
\end{figure*}

\subsection{Multiuser Case}

In this subsection, we consider the multiuser case with $K=3$. The resolutions of the IRS with $B=1,2,3$ and $\infty$ are considered in the following numerical results, where $B=\infty$ stands for the continuous phase shifts for the IRS. Without loss of generality, we assume the SNR requirements are the same for all $K$ users, namely, $\gamma_k=\gamma, \forall k$.

In Fig. 6, we plot some realizations of the signal received at the users with QPSK, 8-PSK, and 16-QAM constellations, respectively. We can intuitively observe that the SER performance deteriorates with the increase of the constellation order under the same SNR requirements. The underlying reason is that the average CIR shrinks as the constellation order increases. Thus, for the symbol-level precoding, the constellation has a significant impact on the user's SER performance.

\begin{figure}
\begin{center}
\includegraphics[width=7.5 cm]{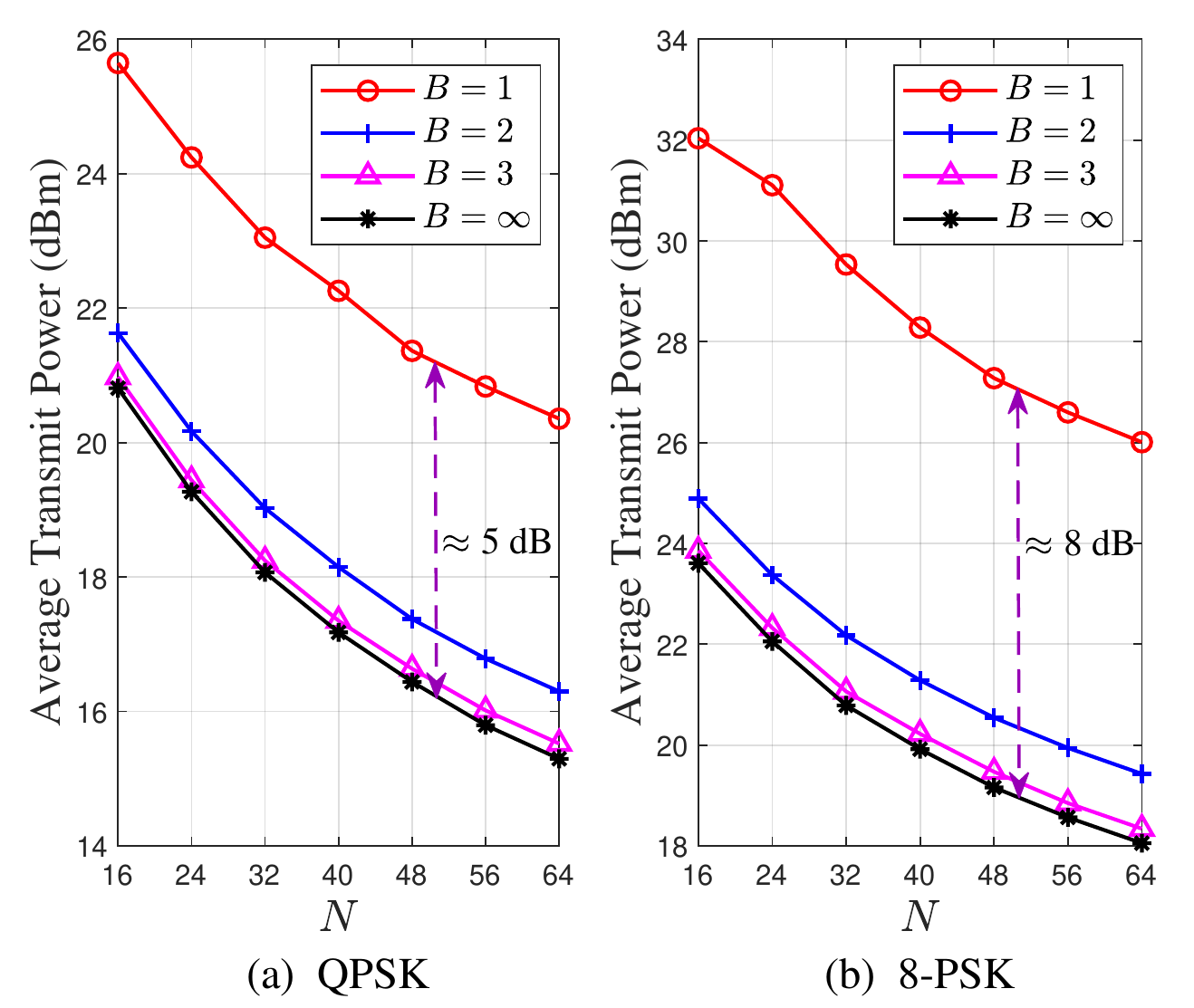}
\caption{Average transmit power versus the number of reflective elements $N$ ($\gamma=10$ dB, $\delta=0.02$).}
\label{Casimir}
\end{center}
\vspace{-0.8cm}
\end{figure}

In Fig. 7, it is shown that the average transmit power decreases greatly with the number of the reflective elements $N$. Therefore, to achieve a low energy consumption, a large number of the IRS elements is needed in the system. Moreover, for both QPSK and 8-PSK constellations, we can see that the resolution of the phase shifters has a significant impact on the transmit power required by an IRS-aided communication system. With the increase of the resolution, the average transmit power at the BS decreases. Since we can tune the phase shift vector much elaborately such that a lower transmit power can be achieved to satisfy the performance requirements. Notably, compared with the average power obtained in the case of $B=1$, a large performance gain can be achieved when $B=2$ for both QPSK and 8-PSK constellations. However, when $B=3$ or higher, the performance gain vanishes and the transmit power approaches to that of in the continuous phase shifts case. Therefore, a tradeoff between the system performance and the hardware implementation should be carefully designed by setting the phase shifters' resolution. In addition, by comparing Fig. 7(a) with (b), we can observe that a larger transmit power is needed for the higher-order constellation. The performance gap between $B=1$ and $B=\infty$ is larger for the 8-PSK constellation than that of for the QPSK constellation. Besides, the performance gap between two given IRS resolutions is almost constant for the different number of IRS elements $N$.

\begin{figure}[t]
\centering
\subfigure[]{
\begin{minipage}{7.5 cm}
\centering
\includegraphics[width=7.5 cm]{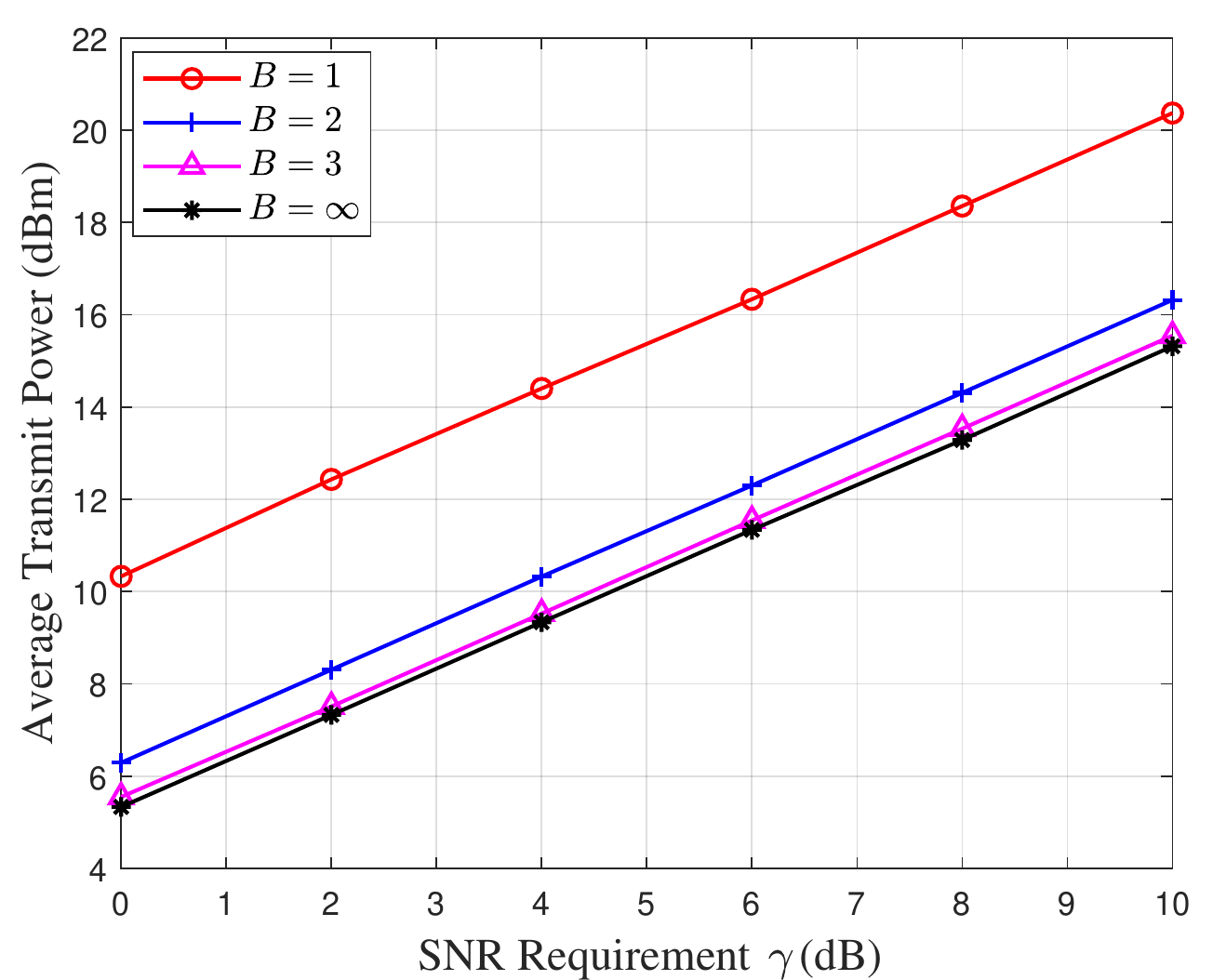}
\end{minipage}%
}
\subfigure[]{
\begin{minipage}{7.5 cm}
\centering
\includegraphics[width=7.5 cm]{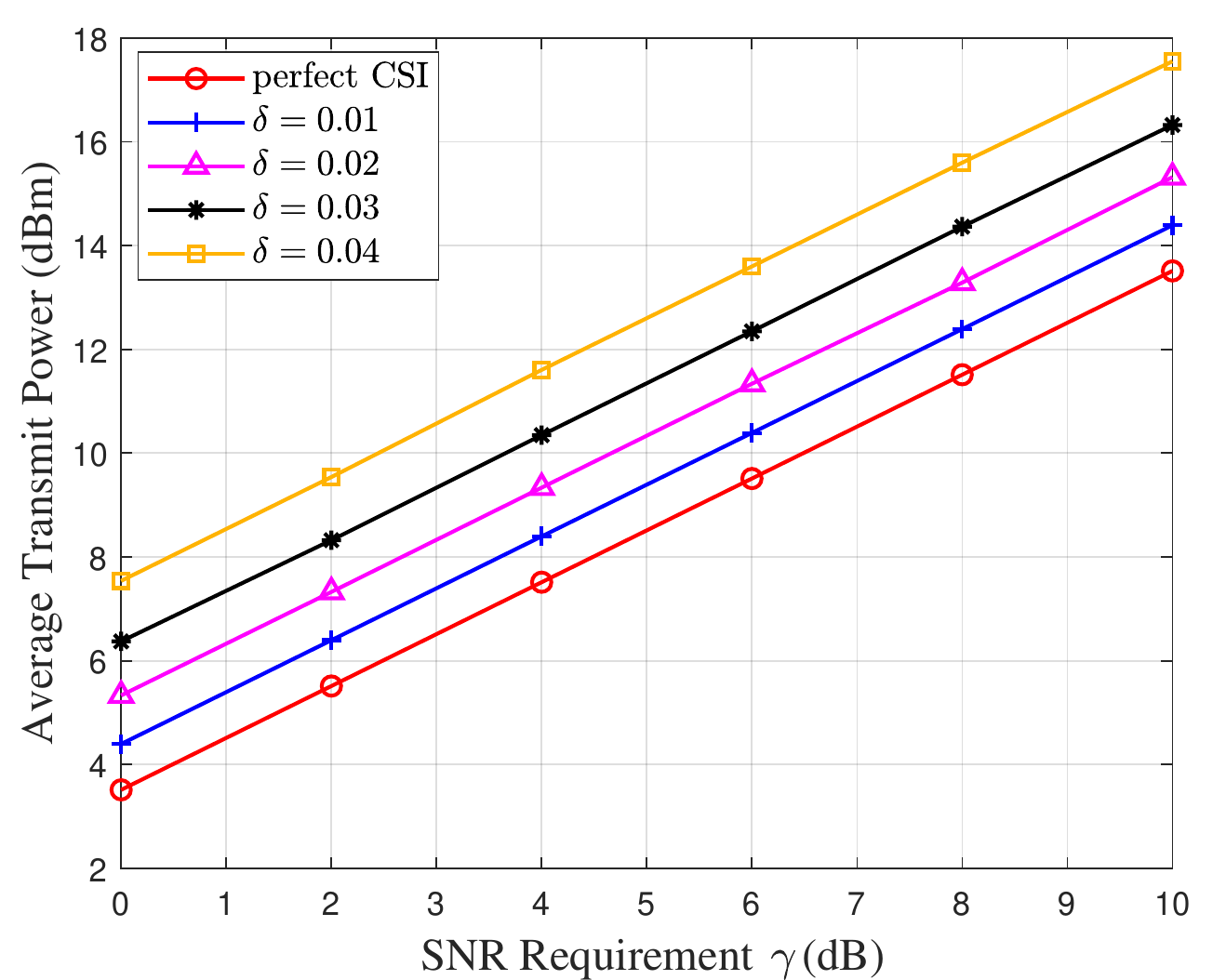}
\end{minipage}%
}
\caption{Average transmit power versus the SNR requirement $\gamma$, (a) $N=64$, $\delta=0.02$; (b) $N=64$, $B=\infty$.}
\vspace{-0.5cm}
\end{figure}

From Fig. 8(a), we can see that the average transmit power increases linearly with the increase of SNR requirements. Actually, for the case of the homogeneous SNR requirements, it can be proved that the transmit power increases by 2 dBm when the SNR requirements increase by 2 dB according to the structure of problem \eqref{D}. Additionally, from Fig. 8(b), it is shown that the transmit power increases with the channel uncertainty level $\delta$ to guarantee the robustness of the system.

Next, let us consider the system performance in terms of SER. As shown in Fig. 9(a), the average SER reduces with the increase of the SNR requirement. Interestingly, we can see that the SER performance degrades with the increase of the IRS resolution. It implies that the extra transmit power due to the low resolution can provide a performance gain in SER. For example, to achieve the specification of $\text{SER}=10^{-3}$, the SNR requirement for $B=2$ is $0.45$ dB higher than that of $B=1$. However, it should be noted that the performance remains almost unchanged when $B\geq2$ with a given SNR requirement. To assess the impact of the channel uncertainty, the average SER versus the SNR requirement with different channel uncertainties is plotted in Fig. 9(b). We can observe that a greater SER performance can be achieved when the channel uncertainty is larger. It is mainly due to the fact that the conservativeness introduced to guarantee the worst-case robustness increases as the channel uncertainty.

\begin{figure}[t]
	\centering
	\subfigure[]{
		\begin{minipage}{7.5 cm}
			\centering
			\includegraphics[width=7.5 cm]{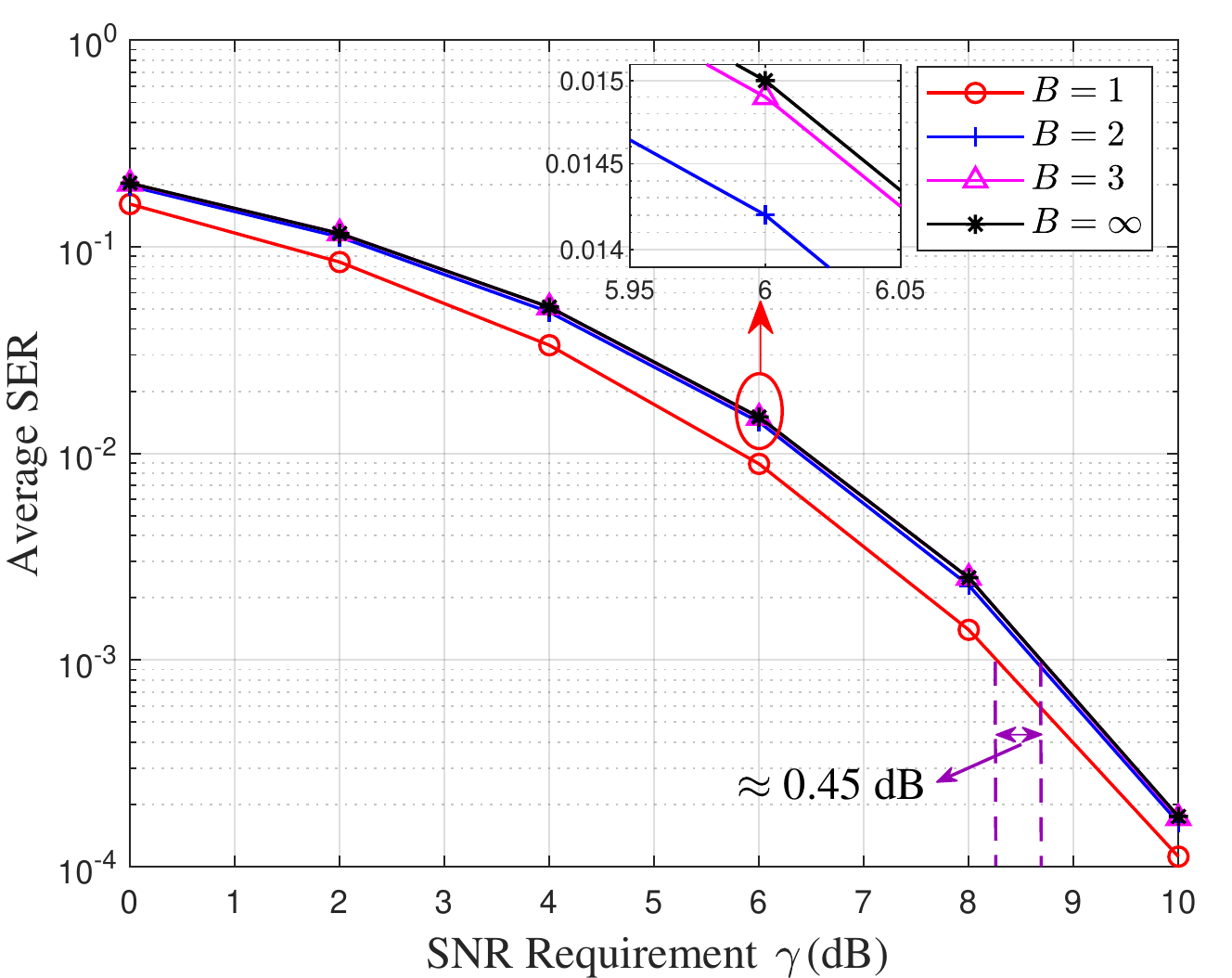}
		\end{minipage}%
	}
	\subfigure[]{
		\begin{minipage}{7.5 cm}
			\centering
			\includegraphics[width=7.5 cm]{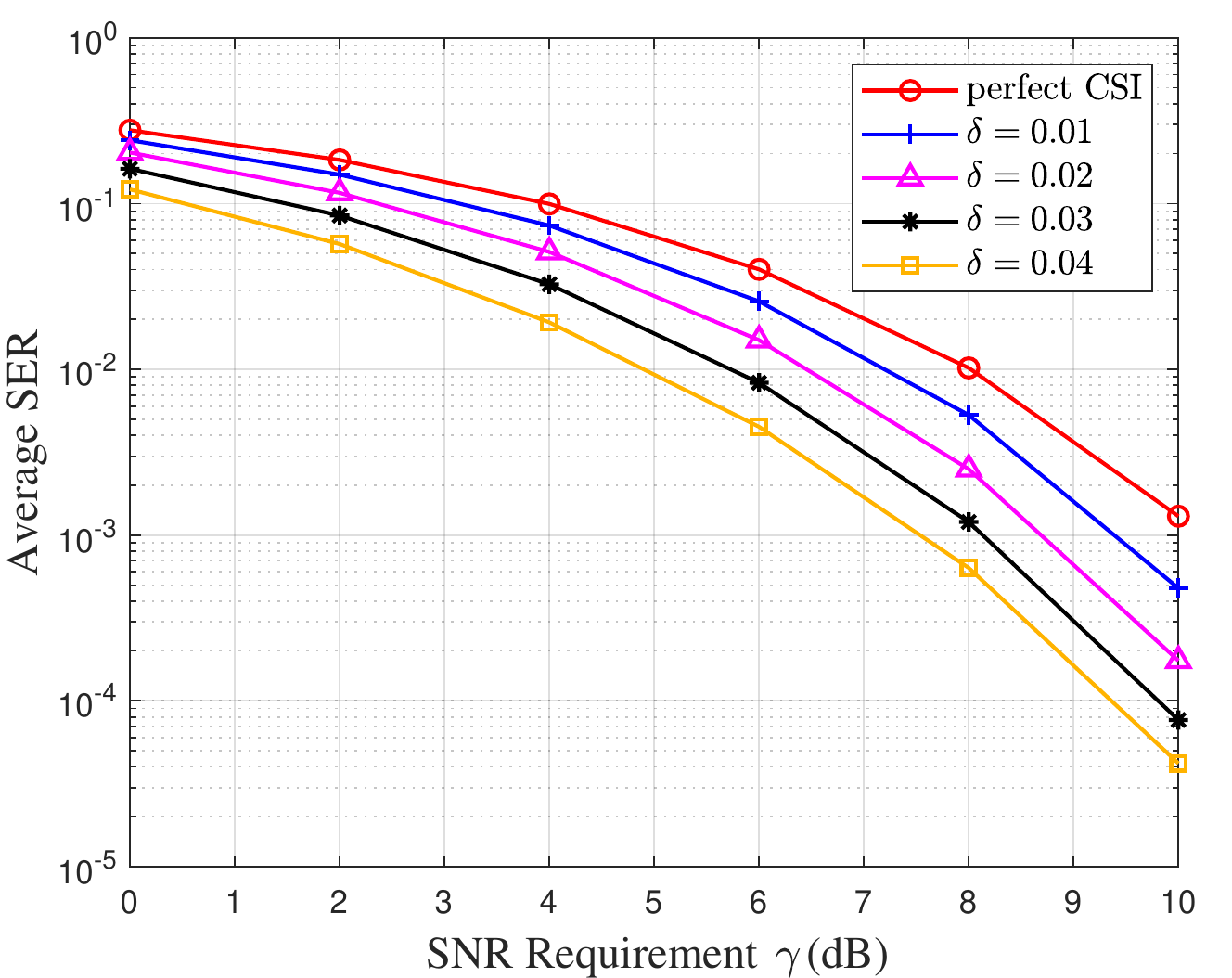}
		\end{minipage}%
	}
	\caption{Average SER versus the SNR requirement $\gamma$, (a) $N=64$, $\delta=0.02$; (b) $N=64$, $B=\infty$.}
	\vspace{-0.4cm}
\end{figure}

\begin{figure}[t]
\begin{center}
\includegraphics[width=7.5 cm]{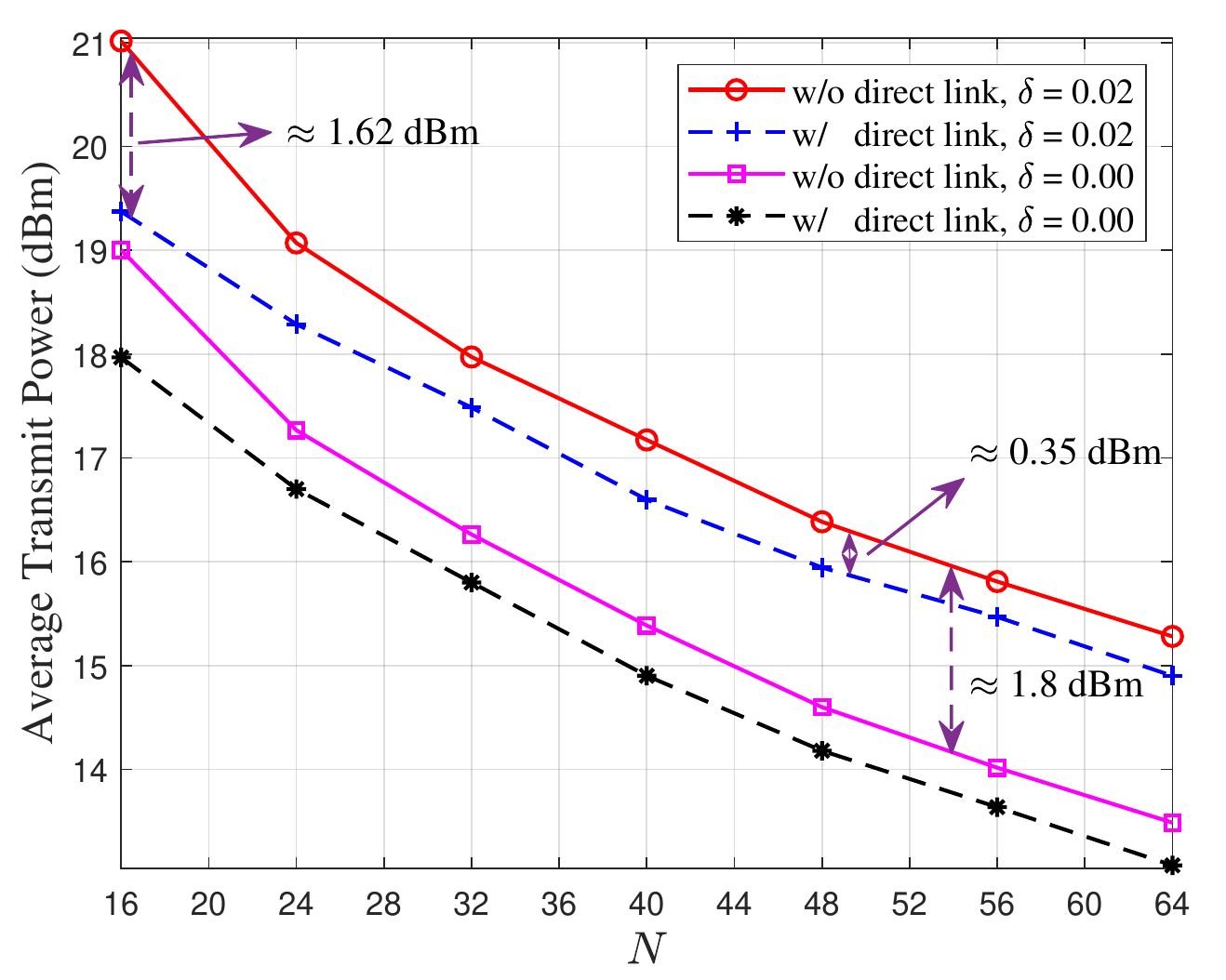}
\caption{Average transmit power versus the number of reflective elements $N$ ($\gamma=10$ dB, $B=\infty$).}
\label{Casimir}
\end{center}
\vspace{-0.5cm}
\end{figure}

Finally, we consider the scenario with direct links between the BS and the users. As shown in Fig. 10, we can observe that there is a performance improvement when the direct links exist. Specifically, when the number of IRS elements is small, such as $N=16$, the existence of the direct links plays a significant role to decrease the transmit power at the BS, and the performance gain of about $1.62$ dBm is obtained. However, with the increased number of the IRS elements, the performance gain introduced by the direct links decreases and then maintains almost constant such as $0.35$ dBm in the case of channel uncertainty level $\delta=0.02$. Additionally, we can find that eliminating the uncertainty of the reflective channel has a more significant impact on the average transmit power compared with exploiting the direct links when $N$ is large.

\section{Conclusion}

In this paper, we investigated a worst-case robust beamforming design with symbol-level precoding in a downlink IRS-aided MISO communication system. We aimed to minimize the transmit power by jointly designing the symbol-level precoding and the reflective vectors at the BS and the IRS, respectively. Two methods, namely an SDR-based method and an AO-based method, were proposed to solve the problem in the single-user system. For the multiuser case, an efficient algorithm based on the AO method and PGD method was developed. The simulation results verified the effectiveness of the algorithm that we proposed. Besides, the results demonstrated the impact of the IRS resolution on the transmit power and SER performance. It is also shown that the direct links can provide a performance gain.

\appendices
\section{Proof of Lemma 1}
\begin{IEEEproof}
For the BPSK constellation, $\bf{D}$ is a diagonal matrix and defined as
\begin{align}
{\bf{D}}=2
\begin{bmatrix}
{\bf{I}}_N &\bf{0}\\
\bf{0}  &-{\bf{I}}_N
\end{bmatrix}\in\mathbb{R}^{2N\times2N}.
\end{align}
Then, according to the definition of $\bar{\bf{H}}$, we have
\begin{align}
\bar{\bf{H}}=\mathcal{T}(\hat{\bf{H}})=
\begin{bmatrix}
\Re(\hat{\bf{H}}) &-\Im(\hat{\bf{H}})\\
\Im(\hat{\bf{H}}) &\Re(\hat{\bf{H}})
\end{bmatrix}\in\mathbb{R}^{2N\times 2M}.
\end{align}
For notational simplicity, let us define ${\bf{P}}\triangleq\Re(\hat{\bf{H}})\in\mathbb{R}^{N\times M}$ and ${\bf{Q}}\triangleq\Im(\hat{\bf{H}})\in\mathbb{R}^{N\times M}$. Then, we can obtain
\begin{align}
&{\bf{W}}\triangleq\bm{{\rm D}}\bar{\rm\mathbf{H}}\bar{\rm\mathbf{H}}^T\bm{{\rm D}}^T=4
\begin{bmatrix}
{\bf{P}\bf{P}}^T+{\bf{Q}\bf{Q}}^T  &{\bf{Q}\bf{P}}^T-{\bf{P}\bf{Q}}^T\\
{\bf{P}\bf{Q}}^T-{\bf{Q}\bf{P}}^T  &{\bf{P}\bf{P}}^T+{\bf{Q}\bf{Q}}^T
\end{bmatrix}\in\mathbb{R}^{2N\times 2N}.
\end{align}

It should be noticed that ${\bf{Q}\bf{P}}^T-{\bf{P}\bf{Q}}^T$ and ${\bf{P}\bf{Q}}^T-{\bf{Q}\bf{P}}^T$ are antisymmetric matrices. In consequence, we can draw the following conclusion
\begin{align}
\text{tr}\left({\bf{W}}
\begin{bmatrix}
{\bf{x}}_1\\
{\bf{x}}_2
\end{bmatrix}
[{\bf{x}}_1^T ~{\bf{x}}_2^T]\right)=
\text{tr}\left({\bf{W}}
\begin{bmatrix}
{-\bf{x}}_2\\
{\bf{x}}_1
\end{bmatrix}
[-{\bf{x}}_2^T ~{\bf{x}}_1^T]\right),
\end{align}
where ${\bf{x}}_1\in\mathbb{R}^N$ and ${\bf{x}}_2\in\mathbb{R}^N$. Moreover, the two vectors $[{\bf{x}}_1^T ~{\bf{x}}_2^T]$ and $[-{\bf{x}}_2^T ~{\bf{x}}_1^T]$ are linearly independent except that ${\bf{x}}_1$ and ${\bf{x}}_2$ are zero vectors.

Next, let us turn to problem \eqref{S6}. Note that problem \eqref{S6} is always feasible, e.g., ${\bf{\Theta}}=\frac{1}{2}{\bf{I}}_{2N}$, Since the constraint set is compact, there exists an optimal solution to \eqref{S6}, denoted by $\tilde{\bm{\Theta}}^*$. Due to the constraint \eqref{S6c}, we can express the optimal solution as
\begin{align}
\tilde{{\bf{\Theta}}}^*=\sum_{i=1}^{2N}\lambda_i\tilde{\bm{\theta}}_i^*(\tilde{\bm{\theta}}_i^*)^T,
\end{align}
where $\lambda_i\geq 0$ is the eigenvalue of $\tilde{\bf{\Theta}}^*$, and $\tilde{\bm{\theta}}_i^*$ is the eigenvector corresponding to $\lambda_i$.

Denote $\tilde{\bm{\theta}}_i^*=[{\tilde{\bm{\theta}}_{i1}}^T,~{\tilde{\bm{\theta}}_{i2}}^T]^T,\forall i,\lambda_i>0$, where $\tilde{\bm{\theta}}_{i1}\in\mathbb{R}^N$ and $\tilde{\bm{\theta}}_{i2}\in\mathbb{R}^N$. Then, we can construct a vector $\tilde{\bm{\theta}}_i^\star=[-{\tilde{\bm{\theta}}_{i2}}^T,~{\tilde{\bm{\theta}}_{i1}}^T]^T$ which is linearly independent of $\tilde{\bm{\theta}}_i^*$. Based on the structure of $\mathbf{B}_n$, we have $\text{tr}(\mathbf{B}_n\mathbf{B}_n^T\tilde{\bm{\theta}}_i^*(\tilde{\bm{\theta}}_i^*)^T)=\text{tr}(\mathbf{B}_n\mathbf{B}_n^T\tilde{\bm{\theta}}_i^\star(\tilde{\bm{\theta}}_i^\star)^T), \forall n$. Therefore, according to (67) and (68), we can construct another optimal solution
\begin{align}
\tilde{\bf{\Theta}}^{\star}=\lambda_i\left(\alpha\tilde{\bm{\theta}}_i^*(\tilde{\bm{\theta}}_i^*)^T+(1-\alpha)\tilde{\bm{\theta}}_i^\star(\tilde{\bm{\theta}}_i^\star)^T\right)+\sum_{\ell=1,\ell\neq i}^{2N}\lambda_{\ell}\tilde{\bm{\theta}}_\ell^*(\tilde{\bm{\theta}}_\ell^*)^T
\end{align}
such that $\text{tr}({\bf{W}}\tilde{\bm{\Theta}}^*)=\text{tr}({\bf{W}}\tilde{\bm{\Theta}}^\star)$ for any $\alpha\in[0, 1)$. By adjusting $\alpha$, an infinite number of the optimal solutions to problem \eqref{S6} can be obtained. Based on the above analysis, if there is an optimal solution with two eigenvectors like $\tilde{\bm{\theta}}_i^*$ and $\tilde{\bm{\theta}}_i^\star$, then we can always obtain a lower-rank optimal solution. This completes the proof of Lemma 1.
\end{IEEEproof}

\begin{spacing}{1.1}
\small
\bibliographystyle{IEEEtran}
\bibliography{reference}{}	
\end{spacing}
	
\end{document}